\documentclass[prb,aps,floatfix,amsmath,amssymb,superscriptaddress]{revtex4}

\usepackage{mathtools}
\usepackage{amsfonts}
\usepackage{amssymb}

\usepackage{graphicx}
\usepackage[all]{xy}

\usepackage{amsthm}

\newcommand{\fm}{{\cal F}}
\newcommand{\gm}{{\cal G}}
\newcommand{\um}{{\cal A}}
\newcommand{\vm}{{\cal B}}

\newcommand{\tf}{\tilde f}
\newcommand{\tF}{\tilde F}

\newtheorem{theorem}{Theorem}[section]
\newtheorem{lemma}[theorem]{Lemma}
\newtheorem{definition}[theorem]{Definition}

\newcommand{\be}{\begin{equation}}
\newcommand{\ee}{\end{equation}}

\begin{document}
\title{Notes on Some Questions in Mathematical Physics and Quantum Information}
\author{M. B. Hastings}
\affiliation{Station Q, Microsoft Research, Santa Barbara, CA 93106-6105, USA}
\affiliation{Quantum Architectures and Computation Group, Microsoft Research, Redmond, WA 98052, USA}
\begin{abstract}
This is a set of notes on some unrelated topics in mathematical physics, at varying levels of detail.  First, I consider certain questions relating to the decay
of correlation functions in matrix product states, in particular those generated by quantum expanders.  This is discussed in relation to recent results of Brandao and Horodecki on area laws on systems with exponentially decaying correlation function\cite{areaexp}.  Second, I consider some difficulties in trying to construct 
a tensor product state (or PEPS) describing a two-dimensional fermionic system with non-vanishing Hall conductance.  Third, I present some relations between the theory of almost commuting matrices and that of vector bundles, making the connection between the classifications more explicit.
Fourth, I present an open question about quantum channels, and some partial results.
\end{abstract}
\maketitle

This is a collection of notes on some topics in mathematical physics.  The topics are not related to each other.  While some of these results could be turned into a paper, others are partial or are not sufficiently important to justify a detailed presentation.  The intent then is to present these notes in an arxiv-only form in the hope that they might be useful to someone.
Likely much of the referencing is incomplete and there are quite a few details left out.
I thank T. Loring, F. Brandao, J. Yard, A. Harrow, G. Smith, P. Shor, Z. Wang, M. Freedman, N. Read and many others for useful discussions.

\part{Decay of Correlation Functions in Matrix Product States}
\section{General Bounds}
In this part we consider correlation functions in matrix product states.  The goal is to note certain tightened bounds on the correlation functions in specific examples, then note a tension between these results and recent results on area laws, and finally to resolve the apparent contradiction.

Consider a one-dimensional spin system of $N$ sites.  Sites are labelled by integers, $1 \leq i \leq N$.
On each site we have a $d$-dimensional Hilbert space.  The wavefunction $\Psi$ of a matrix product state takes the form
\be
\Psi(s_1,s_2,...,s_N)=A^{(1)}(s_1) A^{(2)}(s_2) A^{(3)}(s_3) ... A^{(N-1)}(s_{N-1}) A^{(N)}(s_N),
\ee
where each $s_i$ labels the state on $s_i$ with $0 \leq i \leq d-1$ in some basis.  Each $A^{(i)}(s_i)$ represents some matrix; for each $i$ there are $d$ such matrices, labelled by different choices of $s_i$.  $v(s_1)$ and $w(s_N)$ are vectors.  The expression
is to be interpreted as a product of matrices.  The column dimension of $A^{(i)}$ must match the row dimension of $A^{(i+1)}$.  The matrix $A^{(1)}$ has row dimension $1$ while the matrix $A^{(N)}$ has column dimension $1$, so that the product above is a $1$-by-$1$ matrix, which is regarded as a scalar giving the amplitude 
$\Psi(s_1,s_2,...,s_N)$.
A useful review of such states is in Ref.~\onlinecite{mporev}.  One of the earliest examples of such states is the AKLT state\cite{aklt}.  The general form was called {\it finitely correlated states}, when it was introduct in Ref.~\onlinecite{fnw}.

Suppose that all of the matrices, other than $A^{(1)}$ and $A^{(N)}$ are the same, so that $A^{(i)}=A$ for some matrix $A$ for all $2 \leq i \leq N-1$.  Let $A$ be a $k$-by-$k$ matrix for some $k$.
Then, it is useful to introduce the following {\it transfer matrix}.  The transfer matrix is a linear operator ${\cal E}$ which acts on $k$- by-$k$ matrices $\rho$ as follows:
\be
{\cal E}(\rho)=\sum_s A^\dagger(s) \rho A(s).
\ee
Assume that ${\cal E}$ is diagonalizable.
Let us normalize the matrices $A$ by multiplying by a scalar (this normalization can be absorbed into a normalization of the matrices $A^{(1)},A^{(N)}$ so that the state $\Psi$ still has $|\Psi|=1$) so that the largest eigenvalue (largest in absolute value) of ${\cal E}$ is equal to
$1$.  Assume that there is only $1$ eigenvalue equal to $1$ and that all other eigenvalues are bounded in absolute value by $\lambda$ for some $\lambda<1$.
Then, in Ref.~\onlinecite{fnw} it is shown that correlation functions in this state decay exponentially.
That is, given an operator $A$ supported on some interval of sites $[P,Q]$ and another operator $B$ supported on some other interval of sites $[R,S]$ with $1 \leq P \leq Q < R \leq S \leq N$,
we have that
\be
|\langle \Psi,A B \Psi \rangle - \langle \Psi, A \Psi \rangle \langle \Psi, B \Psi \rangle | \leq c \Vert A \Vert \Vert B \Vert \lambda^{R-Q},
\ee
where $\Vert ... \Vert$ denotes the operator norm.
The constant $c$ however may depend upon $k$.

This dependence on $k$ is unfortunate, as it suggests that for large $k$ we may not see the exponential decay until $R-Q$ is quite large.
However, we next show that in many specific cases an exponential decay can be obtained with a prefactor that is more tightly bounded.  See Eq.~(\ref{nopref}) for the bound that can be proven under some assumptions on the matrices $A(i)$ and on the support of $A,B$.
These results likely appear elsewhere, but I do not know a specific reference.

\section{Manifestly Hermitian Transfer Matrix}
Note that we make the replacements
\be
A^{(i)} \rightarrow A^{(i)} X \quad \; \quad A^{(i+1)} \rightarrow X^{-1} A^{(i)}
\ee
for any invertible matrix $X$, then this leaves the matrix product state unchanged.
It is conventional in the literature to exploit this freedom to bring the matrices $A$ into a certain canonical form\cite{mporev}.  This freedom is sometimes called a gauge freedom.
Note that ${\cal E}(\rho)$ is a completely positive map; this canonical form corresponds to making ${\cal E}$ a trace-preserving completely positive map; that is, ${\cal E}$ is a quantum channel.

However, such a canonical choice need not be made.  Suppose that the Hilbert space dimension $d$ is even and suppose
that
the matrices $A(i)$ have the property that
\be
A(i+d/2 \mod d)=A(i)^\dagger
\ee
for some gauge choice.  We will say then that ${\cal E}$ is {\it manifestly Hermitian}.  Note that this implies that ${\cal E}$ is Hermitian, when regarded as a linear operator acting on $\rho$ with $\rho$ regarded as a vector in a $k^2$-dimensional space, so that
\be
{\rm tr}\Bigl(\rho^\dagger {\cal E}(\sigma)\Bigr)={\rm tr}\Bigl(({\cal E}(\rho))^\dagger \sigma\Bigr).
\ee

In this case, we will show tighter bounds on the prefactor in the correlation functions above if the operators $A$ and $B$ are supported sufficiently far from the edges of the chain; that is, if $P-1$ and $N-S$ are sufficiently large.  In the next section, we give some specific examples of such ${\cal E}$.

So, for the rest of this section, we assume that ${\cal E}$ is manifestly Hermitian, has a unique eigenvalue equal to $1$, and that all other eigenvalues of ${\cal E}$ are bounded in absolute value by $\lambda$ for some $\lambda<1$.
Let $\Lambda$ be the eigenvector of ${\cal E}$ with eigenvalue $1$, normalized so that
${\rm tr}(\Lambda^\dagger \Lambda)=1$.
Consider the expectation value $\langle \Psi, A B \Psi \rangle$.  Write this as
\be
{\rm tr}(\Lambda_B^\dagger {\cal E}^{R-Q}(\Lambda_A)),
\ee
where $\Lambda_A$ is the matrix on the bond variables connecting spins $Q$ to $Q+1$ given by
summing over spins $s_1,...,s_Q$ and $\Lambda_B^\dagger$ is the matrix on the bond variables connecting spins $R-1,R$ given by summing over spins $s_R,...,s_N$.
Formally, $\Lambda_A$ is given by
\begin{eqnarray}
&&\Lambda_A=\sum_{s_1,...,s_R} \langle s_1,...,s_R| A| s_1,...,s_R \rangle
A^{(R)}(s_R)^\dagger ... A^{(1)}(s_1)^\dagger A^{(1)}(s_1)  ... A^{(R)}(s_R).
\end{eqnarray}

In what follows, $...$ will denote various corrections that tend to zero for $P$ and $N-S$ both taken large.
Let $\Lambda_N^\dagger=\sum_{s_N} A^{(N)}(s_N) A^{(N)}(s_N)^\dagger$, so that $\Lambda_N$ is the matrix on the bond variables conecting spins $N-1,N$ after summing out spin $N$.
Note that
\begin{eqnarray}
\label{Aval}
\langle \Psi,A \Psi \rangle&=&{\rm tr}(\Lambda_N^\dagger {\cal E}^{N-P}(\Lambda_A)) \\ \nonumber
&=&
{\rm tr}(\Lambda^\dagger \Lambda_A)+...,
\end{eqnarray}
and similarly
\be
\label{Bval}
\langle \Psi,B \Psi \rangle={\rm tr}(\Lambda_B^\dagger \Lambda)+...
\ee
We claim that
\be
\label{Abnd}
{\rm tr}(\Lambda_A^\dagger \Lambda_A) \leq \Vert A \Vert^2+...,
\ee
and
\be
\label{Bbnd}
{\rm tr}(\Lambda_B^\dagger \Lambda_B) \leq \Vert B \Vert^2+....
\ee
Once this is shown, it will follow from Eqs.(\ref{Aval},\ref{Bval},\ref{Abnd},\ref{Bbnd}) and Cauchy-Schwartz that
\be
\label{nopref}
|\langle \Psi,A B \Psi \rangle - \langle \Psi, A \Psi \rangle \langle \Psi, B \Psi \rangle | \leq \Vert A \Vert \Vert B \Vert \lambda^{R-Q}+....,
\ee
giving the desired bound.  We will use the fact that ${\cal E}$ is manifestly Hermitian to show Eqs.~(\ref{Abnd},\ref{Bbnd}), while we use the fact that ${\cal E}$ is Hermitian and the assumptions on the spectrum of ${\cal E}$ to show Eq.~(\ref{nopref}) from Eqs.(\ref{Aval},\ref{Bval},\ref{Abnd},\ref{Bbnd}).

To show Eq.~(\ref{Abnd}), let $A_{ref}$ be a ``reflected" version of $A$. The basic idea is to define $A_{ref}$ by reflecting $A$ about the bond from $Q$ to $Q+1$, while also making a basis transformation in such a way that
\be
{\rm tr}(\Lambda_A^\dagger \Lambda_A)=\langle \Psi, A_{ref} A \Psi \rangle +...  \leq \Vert O_A \Vert^2.
\ee
The operator $A_{ref}$ will be supported on sites $Q+1,...,Q+1+(Q-P)$.
To define $A_{ref}$, decompose $A$ as a sum of tensor products of operators on sites $P,P+1,...,Q$:
\be
A=\sum_{o_P,...,o_Q} \phi(o_p,...,o_q) O_P(o_P) \otimes ... O_Q(o_Q),
\ee
where $\phi(o_p,...,o_q)$ is a scalar, where
each index $o_i$ ranges over $d^2$ possible values and where
$O_i(o_i)$ is a basis for operators on site $i$.
Then, define
\be
A_{ref}=\sum_{o_P,...,o_Q} \phi(o_p,...,o_q) O^{ref}_{Q+1+(Q-P)}(o_P) \otimes ... O^{ref}_{Q+1}(o_Q),
\ee
where $O^{ref}_i(o_i)$ is another complete basis of operators on site $o_i$ and where $O^{ref}_{Q+1+j}(o)$ is obtained from
$O_{Q-j}(o)$ by making a basis transformation shifting the state $|n\rangle$ to $|n+d/2 \mod d\rangle$.
The derivation of Eq.~(\ref{Bbnd}) is similar.

\section{Quantum Expanders and Relation With Area Law Results}
Such manifestly Hermitian sets of matrices $A(i)$ were considered in Ref.~\onlinecite{rugqe}.  There, a probablistic construction was considered with the matrices $A(i)$ proportional to randomly chosen unitaries subsect to the constraint of being manifestly Hermitian.  It was shown that this gave a family of examples with
diverging matrix size (and hence diverging entanglement entropy since $\Lambda$ was proportional to the identity matrix), constant $d$, and $\lambda$ bounded by some quantity strictly less than $1$.  Such an ${\cal E}$ is often referred to as a quantum expander\cite{qexp1,qexp2}.
Thus, this family has exponentially decaying correlation functions (with a prefactor that is bounded, independent of matrix size).
The above result presents a seeming contradiction with the recent work of Ref.~\onlinecite{areaexp}, where it was shown that exponential correlation function decay in one dimension implies an area law, giving a bound on entanglement entropy depending only upon the correlation decay and upon $d$.

Further, the calculation of Ref.~\onlinecite{areaexp} does not rely on correlation functions near the edge of the chain; i.e., in all cases, the operators in the correlation function are not supported on sites near $1$ or $N$.

The resolution of the paradox is that in Ref.~\onlinecite{areaexp} uses correlation functions of pairs of operators $A,B$ with $A$ supported on some set of sites such as $[P,Q]$ above and $B$ supported on some set of sites both to the left and right of $A$.  That is, $B$ will be supported on the union of two intervals $[R_1,S_1]$ and $[R_2,S_2]$ with $R_1\leq S_1 < P \leq Q < R_2 < S_2$.
This sitation can be mapped into the situation where $B$ is supported only one one interval which is to the right of $A$ considered above if we {\it fold} the chain.  We define a new chain, whose leftmost site $1$  is identitied with the interval $[P,Q]$ in the original chain.  Site $2$ of the new chain is identified with the pair of sites $P-1$ and $Q+1$ of the old chain and in general site $i$ of the new chain is identified with the pair of sites $P-i+1$ and $Q+i-1$.
This new chain also has a matrix product state description, and again has a manifestly Hermitian transfer matrix ${\cal E}$ with a gap in its eigenspectrum, but now the correlator $A$ is supported near the end of the chain and so the bound above on correlation functions does not apply and we only have the bound of Ref.~\onlinecite{fnw} which includes a prefactor which depends upon the matrix $\Lambda$.

\section{Classical Analogue}
It is interesting then to ask how the correlation function actually does decay in these matrix product states where the $A(i)$ are randomly chosen unitaries and $B$ is supported on a pair of intervals.
We conjecture that the decay is in a power law in ${\rm dist}(A,B)={\rm min}(P-S_1,R_2-Q)$.

We now present a heuristic argument for a very slow power law decay in a classical analogue of this model.  We emphasize that this classical analogue is a distinct problem, but it seems to have some similar properties.

In this classical model, we consider a particle doing a random walk on an expander graph with $V$ vertices.  The graph will have a girth $g$ of order $\log(V)$.  Time is discrete, and at each time step the particle randomly jumps to a vertex which neighbors its current vertex.  We write $x(\tau)$ to denote the vertex $x$ as a function of time $\tau$.

Suppose $x(0)$ is chosen to be some fixed vertex $v_0$.
In this case, $x(1)$ will be correlated with $x(\tau)$ up to $\tau$ of order $\log(V)$.  To see this, let the graph have degree $d$.  Let $f(v)$ be a function from vertices to $\{-1,1\}$; let $w$ be one arbitrarily chosen neighbor of $v_0$, and let $f(w)=1$ and let $f(v)=-1$ for all other vertices $v \neq w$.  Then, let $g(v)$ be a function that is equal to $1$ for any vertex $v$ with distance less than $g/2$ of $v_0$ such that the shortest path from $v_0$ to $v$ goes through $w$ (because we have taken the distance less than $g/2$, the shortest path is unique) and $g(v)=-1$ for all other vertices $v$.  Then,
the correlation function
\be
\label{ccf}
\overline{f(x(1)) g(x(\tau))}-\overline{f(x(1))} \; \overline{g(x(\tau))}
\ee
is of order unity for $\tau$ smaller than of order $\log(V)$, where the overline denotes the average over random walks.

Suppose instead that $x(0)$ is chosen uniformly at random.  In this case, it seems that there are no two pairs of functions $f(v),g(v)$ that we can pick such that a correlation function like Eq.~(\ref{ccf}) is not exponentially small in $\tau$ for $\tau$ small compared to the first.  We do not prove this.  However, note that if we pick $f(v)$ to be $1$ on some small set of vertices and $0$ elsewhere, then the correlation function is small because for most choices of $x(0)$ the quantity $f(x(1))$ is zero.  Conversely, if we pick $f(v)$ to be $1$ on a large set of vertices, we find that the random quickly forgets the condition of being on this set.

However, suppose we ask about correlation functions of the form
\be
\label{twotimes}
\overline{f(x(-\tau),x(+\tau)) g(x(0))}-
\overline{f(x(-\tau),x(+\tau))} \; \overline{g(x(0))},
\ee
where $f(v_1,v_2)$ is an arbitrary function of two different vertices.  In this case, the average again is over random walks, and we assume that $x(-\tau)$ is chosen uniformly at random.  This correlation function of two times is the classical analogue of the situation where one considers a correlation of observables in matrix product states with $B$ supported on two distinct intervals.  
It now becomes possible to construct a slowly decaying correlation function, with a decay that we believe is proportional to $\tau^{-1/4}$.
Choose $g(v)$ at random, by choosing $g(v)$ independently for each $v$ uniformly from $\{-1,1\}$.  Then, for each pair $v,w$, we choose $f(v,w)$
to be $-1$ or $1$ depending upon whether the average of $g(x(0))$ over random walks starting at $v$ at $-\tau$ and ending at $w$ at $+\tau$ is negative or positive.  We now analyze this choice of $f,g$.

A key fact about random walks on these graphs is that for times $\tau$ small enough compared to the girth, a random walk starting $x(-\tau)$ travels to some other vertex $x(+\tau)$ such that the distance from $x(-\tau)$ to $x(+\tau)$ is typically roughly proportional to a constant times $\tau$.  Thus, some constant fraction of the steps of the random walk are typically on the shortest path from $x(-\tau)$ to $x(+\tau)$ (we remark  that this statement is restricted to typical random walks with the distance between $x(-\tau)$ and $x(+\tau)$ being proportional to $\tau$; if we instead condition on $x(-\tau)=x(+\tau)$ then most steps are not on the shortest path).  Thus, to analyze the correlation function (\ref{twotimes}), it seems justified to restrict to the case that $x(0)$ is on the shortest path from $x(-\tau)$ to $x(+\tau)$, as this is a significant contribution to the correlation function.  However, assuming that $x(0)$ is on this shortest path, typically $x(0)$ is within a distance of order $\tau^{1/2}$ of the midpoint of this path.  Thus, given $x(-\tau),x(+\tau)$ a distance of order $\tau$ apart from each other, the vertex $x(0)$ is drawn from roughly one of $\tau^{1/2}$ possible vertices.
Hence, the average of $g(x(0))$ over the possible choices of $x(0)$ will be of order $\tau^{-1/4}$ and so choosing $f(x(-\tau),x(+\tau))$ as above gives a correlation function of order $\tau^{-1/4}$.

\part{Tensor Product States with Non-Vanishing Chern Number}
In this part, we consider the problem of trying to construct a tensor product state (often called PEPS\cite{peps}) with a bounded bond dimension that will
describe a state with nonzero Hall conductance.
This problem has been mentioned by various people.  Ref.~\onlinecite{read} constructs such a trial state, but unfortunately only for a gapless parent Hamiltonian.  See also Refs.~\onlinecite{peps1,peps2}.
In this part, we presents some arguments showing the difficulty in constructing such a state (this argument relies, however, on many assumptions which are indicated below).

It is not completely clear how to define the problem precisely in the case of an arbitrary tensor product state, since we need a Hamiltonian to define the Hall conductance.  One possible definition is: we want the PEPS to be the exact ground state of a Hamiltonian
(perhaps the parent Hamiltonian of the PEPS), such that the following properties hold.  First, the Hamiltonian should be charge conserving.  Second, the Hamiltonian should have a bound on the range of the interactions.
These two properties enable us to define a Hamiltonian on the torus with twisted boundary conditions (we use angles $\theta_x,\theta_y$ to denote boundary condition twist on the torus); we write this Hamiltonian as $H(\theta_x,\theta_y)$.  Third, for sufficiently large system sizes, the Hamiltonian on the torus should be gapped for all choices of boundary condition twist.  This enables us to define a projector $P(\theta_x,\theta_y)$ which projects onto the ground state for given boundary angles.  Fourth, this projector $P(\theta_x,\theta_y)$ should define a vector bundle with nonvanishing Chern number corresponding to the nonvanishing Hall conductance.

It is possible that the requirement of a gap for all boundary condition angles should be relaxed to requiring just a gap for $\theta_x=\theta_y=0$.  After all, it is proven\cite{hasmich} that for local Hamiltonians such a gap implies that the deviation of the Hall conductance (as defined by the Kubo formula) from an integer is small.  However, here we will impose this strict requirement of a gap for all $\theta_x,\theta_y$.

\begin{figure}
\label{torusint}
\includegraphics[width=3in]{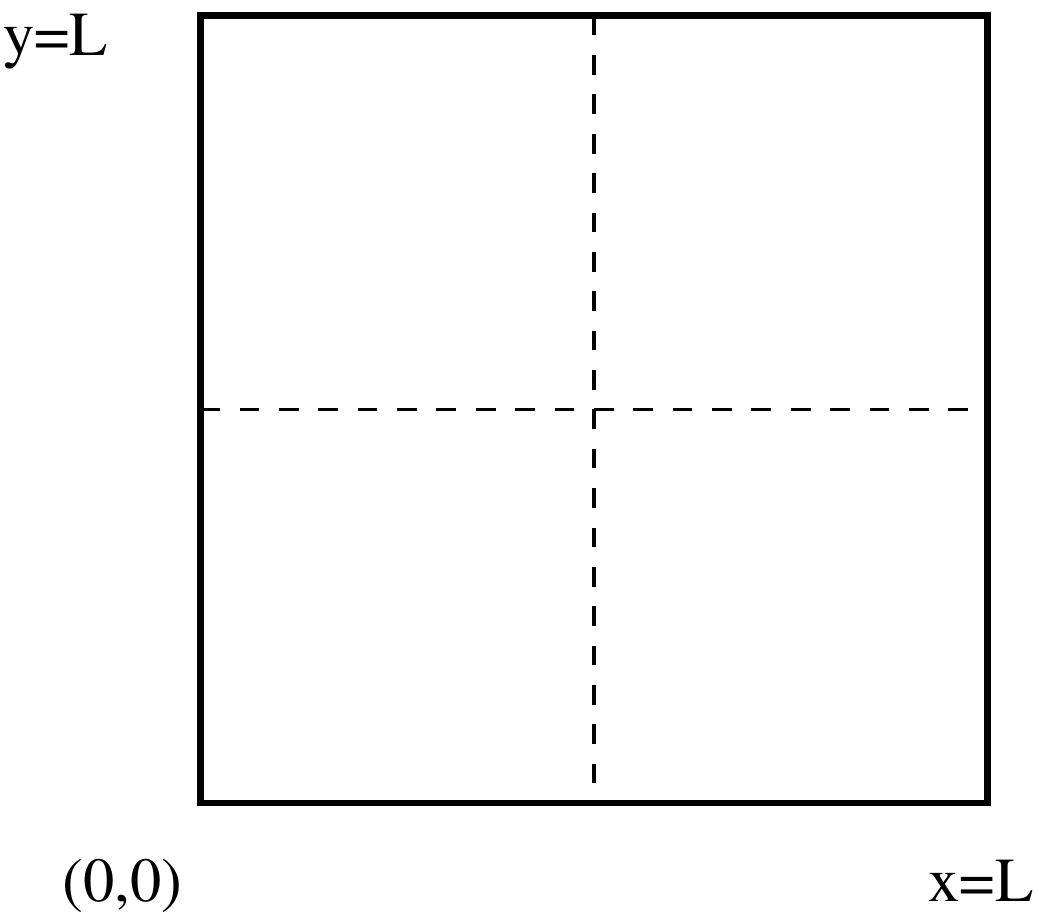}
\end{figure}

For a sketch of the geometry of the system, see Fig.~\ref{torusint}.  This describes a system of linear size $L$ with opposite edges of the square identified.  The boundary twists $\theta_x,\theta_y$ are inserted on the solid vertical and horizontal lines, respectively.  The dashed lines are described below.

We now present some arguments showing the difficulty in constructing such a state.  The difficulty will not be a difficulty in constructing the projector $P(\theta_x,\theta_y)$ but rather a difficulty in constructing a state $\Psi$ as a matrix product state such that $P(0,0)=|\Psi\rangle\langle\Psi|$.  The basic idea is that, under some assumptions noted below, if we can construct a tensor product state $\Psi$ which is a ground state of $H(0,0)$, then we can also construct a family of tensor product states $\Psi(\theta_x,\theta_y)$ which are ground states of $H(\theta_x,\theta_y)$.  Further, these states $\Psi(\theta_x,\theta_y)$ will depend smoothly and periodically on $\theta_x,\theta_y$.  However, this will give a contradiction: constructing such a family $\Psi(\theta_x,\theta_y)$ trivializes the bundle defined by $P(\theta_x,\theta_y)$ which is not possible given the nontrivial Chern number.

The first assumption is that $\Psi$ is given as a translationally-invariant injective tensor product state (see Ref.~\onlinecite{dpg} for definitions of injective; here we assume also that not only is the state $\Psi$ translationally invariant, but that the tensors defining are also translationally invariant).
Then, as shown in Ref.~\onlinecite{dpg}, there is a way of writing the tensors defining the state such that a symmetry transformation on the physical spins is realized on the bond variables.  This is a constraint on the tensors of the state, shown graphically in Fig.~11 of the arxiv verson of Ref.~\onlinecite{dpg}.
That is, let $Q$ be the total charge of the system, with $Q=\sum_i q_i$, where $q_i$ is a charge operator on each site $i$.  The operator $Q$ commutes with $H$.  Then, under these assumptions, we can write the state $\exp(i q_i \theta) |\Psi\rangle$ by inserting two-index tensors on the bond variables leaving site $i$; that is, these tensors are inserted between the tensor on site $i$ and the tensor on a site neighboring $i$ which is joined to $i$ by a bond variable.  Further, these tensors can be chosen so that the tensor on the bond variable going to the right from site $i$ is the inverse of the tensor on the bond variable going to the left, while the tensor on the bond variable going upwards is the inverse of the tensor on the bond variable going downwards.

Suppose then that $R$ is the region in between the vertical dashed line and the vertical solid line in Fig.~\ref{torusint}, with $R$ to the {\it right} of the vertical dashed line as drawn.
Let $Q_R=\sum_{i \in R} q_i$.  Then $\exp(i Q_R \theta_x) |\Psi\rangle$ can be obtained from $\Psi$ by inserting two-index tensors on the bonds crossing the vertical dashed line and the vertical solid line as all the other rank-two tensors (those on the bonds connecting two sites both in $R$) cancel.
Similarly, let $Q_T$ be the region above the horizontal dashed line and below the horizontal solid line.  Again, $\exp(i Q_T \theta_y) |\Psi \rangle$ can be obtained from $\Psi$ by inserting two-index tensors on the bonds crossing the horizontal lines and $\exp(i Q_R \theta_x + i Q_T \theta)|\Psi\rangle$ can be obtained from $\Psi$ by inserting two-index tensors on bonds crossing horizontal and vertical lines, both dashed and solid.
Now, define $\Psi(\theta_x,\theta_y)$ to be the tensor product state obtained as follows: take the state $\exp(i Q_R \theta_x + i Q_T \theta)|\Psi\rangle$ written as a tensor product state with these two-index tensors inserted, and remove those two-index tensors on bonds crossing the dashed lines (replacing them with the identity on those bonds), keeping only those on the solid lines.

Suppose that $\Psi(\theta_x,\theta_y)$ has the property that the local reduced density matrices of the state on a small region depend only on the tensors in the tensor product state near that region and are insensitive to any change in tensors further away.  If so, then the reduced density matrix of $\Psi(\theta_x,\theta_y)$ will agree with that of $\Psi$ away from both solid lines (because one is far from the bond variables where the two-index tensors have been inserted compared to that state), while near both solid lines it will agree with the reduced density matrix of $\exp(i Q_R \theta_x + i Q_T \theta)|\Psi\rangle$ (as it is far from the bond variables where the two-index tensors have been removed compared to that state).  Similarly, near the vertical solid line and away from the horizontal solid line it will agree with the reduced density matrix of $\exp(i Q_R \theta)|\Psi\rangle$, while near the horizontal solid line and away from the vertical solid line it will agree with the reduced density matrix of $\exp(i Q_T \theta)|\Psi\rangle$.

Thus, under this assumption, if $\Psi(\theta_x,\theta_y)\neq 0$, then $\Psi(\theta_x,\theta_y)$ will be the ground state of $H(\theta_x,\theta_y)$.  (If instead of the reduced density matrix being completely insensitive to a change in further away tensors, but is only exponentially insensitive, then $\Psi(\theta_x,\theta_y)$ will still have a large overlap with the ground state of $H(\theta_x,\theta_y)$).

Next let us assume that $\Psi(\theta_x,\theta_y)$ indeed is non-zero for all $\theta_x,\theta_y$.  This is a trickier assumption to justify, and violations of this assumption present the most likely route to constructing a tensor product state with the desired properties.  To try to justify why we might expect this assumption to hold for many tensor product states, consider evaluating the norm squared of the tensor network.  This is some new two-dimensional tensor network with no external bond variables.  Imagine dividing space into two disjoint regions, $A,B$.  Sum over all sites in $B$, which gives some effective weight on the bond variables connecting $A$ to $B$.  If this weight is independent of changes in the tensor far from the boundary between $A$ and $B$ (i.e., if there is some kind of decay of correlations in the space of bond variables), then we can justify that the norm squared is non-zero for all $\theta_x,\theta_y$, as follows.  Consider first the norm for $\theta_x\neq 0, \theta_y=0$.    Let $A$ be the sites closest to the solid vertical line and $B$ be the sites closest to the dashed vertical line.  We know that if we introduce a change $\exp(i Q_R \theta_x)$ in the wavefunction, this does not change the norm squared.  This change is done by changing the two-index tensors on the bonds crossing both the dashed and solid vertical lines.
The change in the norm due to introducing the two-index tensor on just the solid vertical line or just the dashed vertical line will be some scalar, which we call $z_{solid}$ or $z_{dashed}$.  We want to show that $z_{solid} \neq 0$.  However, using this assumption on correlation decay, and the fact that the norm isn't changed if we introduce tensors on both solid and dashed vertical lines, we have $z_{solid} z_{dashed}=1$, so $z_{solid} \neq 0$.  We can then apply the same argument (using the horizontal lines instead of the vertical lines) to go to $\theta_x \neq 0, \theta_y \neq 0$.

Finally, note that the two-index tensors in the state $\Psi(\theta_x,\theta_y)$ depend smoothly on $\theta_x,\theta_y$ implying that $\Psi(\theta_x,\theta_y)$ depends smoothly on $\theta_x,\theta_y$.  We also want to show that $\Psi(\theta_x,\theta_y)$ depends periodically upon $\theta_x,\theta_y$ with period $2\pi$, but we will assume it does.

Given all these assumptions, then indeed we have constructed a state $\Psi(\theta_x,\theta_y)$ which is a ground state of $H(\theta_x,\theta_y)$ and which depends smoothly on $\theta_x,\theta_y$, giving a contradiction.  There are clearly several assumptions, so there may be a way around it.  However, some of these assumptions will be difficult to evade; for example, violating the injectivity assumption is often associated with topological order, while here we hope to describe a state without topological order (at least, we hope for a unique ground state on the torus).  If one tries to make $\Psi(\theta_x,\theta_y)$ not periodic as a function of $\theta_x,\theta_y$, again this seems difficult; this effect would likely be associated with the presence of fractionalization of charge, which we do not want to happen and which might also violate the assumption of a unique ground state on a torus.  Also, if $\Psi(\theta_x,\theta_y)$ is periodic in $\theta_x,\theta_y$ with a larger period $2\pi m$ for some integer $m>1$, this still gives a contradiction as in this case we can consider the bundle over an $m$-fold cover of the flux torus.

\part{From Almost Commuting Matrices to Vector Bundles, and Back}
In this section,
I define a map from almost commuting unitaries to functions.  These functions map from the torus to a projector on a finite-dimensional space and hence define a vector bundle.  We also define maps in the opposite direction, from these functions back to almost commuting unitaries, and we show that the composition of the two maps, in either order, is close to the identity up to terms which are trivial as defined below.  This may give new examples of sequences of almost commuting unitary matrices which cannot be approximated by exactly commuting unitaries.  Both maps go via an intermediate stage of constructing a set of exactly commuting unitaries and a projector that almost commutes with those unitaries.  I discuss the effects of symmetries.

For a given pair of unitary matrices, $U,V$, consider the distance (measured in operator norm, $\Vert ... \Vert$) to the nearest pair of unitaries $U',V'$ with $[U',V']=0$.
For all $\epsilon>0$, there exist\cite{voiculescu} pairs of finite-dimensional unitary matrices, $U$ and $V$, such that the commutator $\Vert [U,V] \Vert \leq \epsilon$, but such that this distance is at least some constant greater than zero, independent of $\epsilon$.  The classification of such unitaries\cite{loring} is based on an integer index and has much in common with the classification of vector bundles.  The goal of this part is to make the relation more explicit, by constructing a map from almost commuting unitaries to vector bundles, and another map going in the reverse direction, such that the composition of the two maps is approximately equal to the identity up to various ``trivial terms" as defined below.
The basic idea is that one can map from almost commuting matrices to bundles by mapping the matrices to a system of fermions hopping in a tight-bonding model and then considering the flux torus, and that one can map in the other direction by a discretization.

These maps will be defined to map a tuple $U_1,...,U_d$ of unitaries to a vector bundle over the $d$-dimensional torus, and we define them for all $d$.
We construct these maps using some intermediate steps.  In section \ref{definitions}, we give some definitions.  In section \ref{maps1}, we construct a map from a tuple of almost commuting unitaries $U_1,...,U_d$ to a tuple of exactly commuting unitaries $U'_1,...,U'_d$ and a projector $P$ that almost commutes with $U'_i$ for all $i$.  We refer to such a tuple and projector as defining a ``local projector".  We also construct an approximate inverse to this map, up to trivial terms.  The results of this section might be regarded as analogous to Swan's theorem.
Then, in section \ref{maps2}, we construct a map from such local projectors to vector bundles, and also construct an approximate inverse to this up to trivial terms.  Combining the results in sections \ref{maps1},\ref{maps2} gives the maps between almost commuting unitaries and vector bundles.

Many of our results involve some approximation: for example, we start with some unitaries $U_i$ that approximately commute and construct exactly commuting unitaries $U'_i$ and a projector $P$ that approximately commutes with them, with the bound on the commutator $\Vert [U'_i,P] \Vert$ depending on the bound on $\Vert [U_i,U_j] \Vert$.  When mapping a bundle to a local projector, the bound on the commutator $[U'_i,P]$ depends inversely upon the size of the resulting matrices.  One application of the results then is to construct sequences such that the commutators go to zero as the matrix size goes to infinity.

One motivation for our construction is that it enables us to handle certain symmetries of the problem.  For example, in applications in physics, physical symmetries such as time-reversal symmetry translate into certain constraints on the matrices.  These symmetries are preserved under the maps, in a way discussed in section \ref{symmsec}.

Another motivation for our construction is to provide a way of constructing additional explicit examples of unitaries which almost commute but which are far from exactly commuting unitaries.
The example of Voiculescu\cite{voiculescu} is well-known for the case of two almost commuting unitaries.  However, classification results\cite{loring2} have shown the existence of other examples in different symmetry classes and with $d>2$. Of course, for all $d>2$, we can construct a tuple $U_1=U,U_2=V,U_3=...=U_d=I$ from the unitaries $U$ and $V$ above; however these classification results reveal the existence of other obstructions beyond those of this form.
These classification results have only shown existence but have not given the examples explicitly; by using the maps we define, it becomes possible to explicitly construct these if we have an explicit construction of an appropriate vector bundle.

We use physics bra-ket notation for inner and outer products.  Sometimes we will not explicitly write the ket $|...\rangle$ for a vector if it is not needed; i.e., writing $v$ rather than $|v\rangle$.  Note that our inner product is conjugate linear in the second variable rather than the first.
We write the conjugate transpose of an operator $O$ as $O^\dagger$ following physics notation.

All our linear operators will be finite-dimensional matrices.

We make frequent use of computer science style big-O notation.  If we write $A=B+O(...)$, where $A$ and $B$ are matrices, this is shorthand for $\Vert A - B \Vert \leq O(...)$.  The bounds are explicit, but we use the big-O notation to avoid excess constants cluttering the results.

For notational simplicity, we will often write $\vec \theta$ to refer to a tuple of angles $\theta_1,...,\theta_d$, or
$\vec m$ or $\vec n$ to refer to a tuple of integers.

\section{Definitions}
\label{definitions}
\begin{definition}
We define an ``$\epsilon$-soft torus" to be a tuple of unitaries $U_1,...,U_d$, for some given $d$, such that
\be
\Vert [U_i, U_j] \Vert \leq \epsilon
\ee
for all $i,j$.
\end{definition}
We will sometimes write ${\cal U}$ to denote a given $\epsilon$-soft torus; i.e., ${\cal U}$ denotes a tuple of unitaries $U_1,...,U_d$ with the properties above.

\begin{definition}
We define an ``$\epsilon$-local projector" to be unitaries $U_1,...,U_d$ and a projector $P$ such that
\be
[U_i,U_j]=0
\ee
for all $i,j$ and
\be
\Vert [P,U_i] \Vert \leq \epsilon.
\ee

We will sometimes write ${\cal P}$ to denote a given $\epsilon$-local projector; i.e., ${\cal P}$ denotes a tuple of unitaries $U_1,...,U_d$ and a projector $P$.

In general, we will refer to any operator $O$ as ``$\epsilon$-local with respect to unitaries $U_i$" if $\Vert [O,U_i \Vert \leq \epsilon$.
\end{definition}
Note that $U_1,...,U_d$ in the above definition form a $0$-soft torus (i.e., an $\epsilon$-soft torus for $\epsilon=0$).

Sometimes if we do not wish to explicitly give $\epsilon$, we will simply refer to an $\epsilon$-soft torus or an $\epsilon$-local projector as a ``soft torus" or ``local projector", respectively.

\begin{definition}
\label{sum}
Given an $\epsilon$-soft torus ${\cal U}$, corresponding to a tuple of unitaries $U_1,...,U_d$ and an $\epsilon'$-soft torus ${\cal V}$, corresponding to a tuple of unitaries $V_1,...,V_d$, we define their sum
${\cal U}+{\cal V}$ to be the tuple of unitaries $U_1 \oplus V_1,...,U_d \oplus V_d$.  Then, ${\cal U}+{\cal V}$ is a ${\rm max}(\epsilon,\epsilon')$-soft torus.
Here, the $\oplus$ denotes the direct sum of matrices.  It is not necessary that $U_i,V_i$ have the same dimension.

Given an $\epsilon$-local projector ${\cal P}$ corresponding to unitaries $U_1,...,U_d$ and projector $P$ and an $\epsilon'$-local projector ${\cal Q}$ corresponding to unitaries $V_1,...,V_d$ and projector $Q$, we define their sum
${\cal P}+{\cal Q}$ to be the tuple of unitaries $U_1 \oplus V_1,...,U_d \oplus V_d$ and projector $P \oplus Q$.
Then, ${\cal P}+{\cal Q}$ is a ${\rm max}(\epsilon,\epsilon')$-local projector.
\end{definition}

\begin{definition}
\label{trivial}
Define a soft torus ${\cal U}$ to be trivial if it is an $\epsilon$-soft torus for $\epsilon=0$.
Define a local projector ${\cal P}$ to be trivial if the projector $P$ corresponding to ${\cal P}$ exactly commutes with the unitaries $U_i$.
\end{definition}

\begin{definition}
Given two soft tori, ${\cal U}$ and ${\cal V}$, with corresponding tuples of unitaries $U_1,...,U_d$ and $V_1,...,V_d$, define a distance
${\rm dist}({\cal U},{\cal V})$ to be the minimum over all unitaries $Y$ of
\be
{\rm max}_i \Vert Y^\dagger U_i Y - V_i \Vert.
\ee
\end{definition}

We now define a distance between local projectors.
\begin{definition}
\label{LocHamDistDef}
Consider an $\epsilon$-local projector ${\cal P}$ corresponding to unitaries $U_1,...,U_d$ and projector $P$ and an $\epsilon'$-local projector ${\cal Q}$ corresponding to unitaries $V_1,...,V_d$ and projector $Q$,
such that $P$ and $Q$ have the same rank.
Define the distance ${\rm dist}({\cal P},{\cal Q})$ as follows.  This distance is the minimum over all trivial local projectors ${\cal R}$ and ${\cal S}$ satisfying a certain condition $*$ (given in the next paragraph) of a quantity that we call $d({\cal P}+{\cal R},{\cal Q}+{\cal S})$.

Let $p$ and $q$ be the projectors corresponding to ${\cal P}+{\cal R}$ and ${\cal Q}+{\cal S}$.  The condition $*$ is that $p$ and $q$ are matrices of the same size as each other and that they have the same rank as each other.
Let the tuple of unitaries corresponding to ${\cal P}+{\cal R}$ be $u_1,...,u_d$ and let the tuple of unitaries corresponding to ${\cal Q}+{\cal S}$ be $v_1,...,v_d$.
Then, let $d({\cal P}+{\cal R},{\cal Q}+{\cal S})$ be the minimum over all unitaries $Y$ such that $Y^\dagger p Y = q$ of
\be
{\rm max}_i \Vert Y^\dagger u_i Y - v_i \Vert.
\ee
\end{definition}

\begin{definition}
\label{closeto}
Having defined these distances, we say that a map from $\epsilon$-soft tori to $\epsilon'$-soft tori is $\delta$-close to the identity if it maps every $\epsilon$-soft torus ${\cal U}$ to an $\epsilon'$-soft torus ${\cal U}'$ with
${\rm dist}({\cal U},{\cal U}')\leq \delta$.  Similarly, we say that a map from $\epsilon$-local projectors to $\epsilon'$-local projectors is $\delta$-close to the identity if it maps every ${\cal P}$ to a ${\cal P}'$ with ${\rm dist}({\cal P},{\cal P}')\leq \delta)$.
\end{definition}

\section{Map Between Soft Tori and Local Projectors}
\label{maps1}
In this section we define a map $\fm$ from $\epsilon$-soft tori to $\epsilon'$-local projectors (the relation between $\epsilon,\epsilon'$ is given below) and also a map $\gm$ from $\delta$-local projectors to $4\delta^2$-soft tori.

Then we show that $\fm \circ \gm$ and $\gm \circ \fm$ are both close to the identity, in the sense of definition \ref{closeto}.

\subsection{Definition of Map $\gm$}
The map $\gm$ has been considered before in Ref.~\onlinecite{HL1,HL2,HL3} and so we give that one first. 
First we need to define the polar of an invertible matrix.
\begin{definition}
Given an invertible matrix $X$, define ${\rm polar}(X)=X (X^\dagger X)^{-1/2}$.
\end{definition}
Note that ${\rm polar}(X)$ is always a unitary matrix.
If $X$ is a unitary, then ${\rm polar}(X)=X$.  It was shown that in Ref.~\onlinecite{HL3} that if $\Vert X^\dagger X - I \Vert \leq \delta$ and
$\delta \leq 0.6$ then $\Vert {\rm polar}(X)-X\Vert \leq \delta$.
Thus, for small $\delta$, ${\rm polar}(X)$ is a unitary that is a close approximation to $X$.

\begin{definition}
We now define the map $\gm$.
Consider an $\epsilon$-local projector ${\cal P}$ with corresponding unitaries $U_1,...,U_d$ and projector $P$; let $\epsilon<0.6$.
For notational convenience we write
\be
\label{notation}
P=\begin{pmatrix} I & 0 \\ 0 & 0 \end{pmatrix},
\ee
where the matrix above is a block matrix with $I$ being the identity matrix and the first block having the same size as the rank of $P$.
Define $U_i^{11}$ by
\be
P U_i P = \begin{pmatrix} U_i^{11} & 0 \\ 0 & 0 \end{pmatrix}.
\ee
Define $\gm({\cal P})$ to be the $\delta$-soft torus corresponding to unitaries ${\rm polar}(U_1^{11}),...,{\rm polar}(U_d^{11})$.
\end{definition}

\begin{lemma}
\label{mapsto}
Let ${\cal P}$ be a $\delta$-local projector with $\delta \leq 0.6$.  Then $g({\cal P})$ is a $4\delta^2$-soft torus.
\begin{proof}
We need to bound $\Vert  {\rm polar}(U_i^{11}), {\rm polar}(U_j^{11})] \Vert$.
Note that $\Vert U_i^{11} (U_i^{11})^\dagger-I \Vert = \Vert (P U_i P) (P U_i P)^\dagger - P \Vert$.
Using the same block matrix notation as in Eq.~(\ref{notation}), write
\be
U_i=\begin{pmatrix} U_i^{11} & U_i^{12} \\ U_i^{21} & U_i^{22} \end{pmatrix}.
\ee
Since $[U_i,U_j]=0$ we have
$0=[U_i^{11},U_j^{11}]+U_i^{12} U_j^{21}-U_j^{12} U_i^{21}$.  
We have $\Vert U_i^{12} \Vert \leq \delta$ and the same bound for 
$\Vert U_i^{21} \Vert$, so $\Vert[U_i^{11},U_j^{11}]\Vert \leq 2 \delta^2$.

Also, we have $U_i U_i^\dagger=I$ so $U_i^{11} (U_i^{11})^\dagger + U_i^{12} (U_i^{12})^\dagger=I$ so
$\Vert U_i^{11} (U_i^{11})^\dagger - I \Vert \leq \delta^2$.
So, 
\be
\label{closetopolar}
\Vert {\rm polar}(U_i^{11})-U_i^{11} \Vert \leq \delta^2.
\ee

Hence, $\Vert  {\rm polar}(U_i^{11}), {\rm polar}(U_j^{11})] \Vert \leq 4 \delta^2$, which follows from the bounds above and from a triangle inequality.
\end{proof}
\end{lemma}

\subsection{Definition of Map $\fm$}
We now define $\fm$ after some preliminaries.

Consider an $\epsilon$-soft torus ${\cal U}$ with unitaries $U_1,...,U_d$.
Define
\be
X_i=\frac{U_i +U_i^\dagger}{2},
\ee
\be
Y_i=\frac{U_i-U_i^\dagger}{2i}.
\ee
Note that $\Vert X_i \Vert \leq 1, \Vert Y_i \Vert \leq 1$.

Define a POVM (positive operator-valued measure)\cite{POVM} as follows.
Let $\Delta$ be some small positive number chosen later in Eq.~(\ref{deltadef}); this choice is made to minimize certain error bounds later.
Let $F(x)$ be a function from real numbers to real numbers such that $F(x)=0$ for $|x| \geq 1$ and such that $F(x-1)+F(x)+F(x+1)=1$ for $|x| \leq 1$.  Then, for all $x$
\be
\label{sumrule}
\sum_n F(x+n)=1
\ee
where the sum is over integer $n$.  Finally, pick $F(x)$ to be non-negative for all $x$, and such that $F(x)^{1/2}$ is infinitely differentiable.

Define the POVM to be the set of operators $E_{\vec m,\vec n}$, where $m_1,...,m_d,n_1,...,n_d$ are integers with $|n_i| \leq \lceil 1/\Delta \rceil$, where we define
\be
\label{Edef}
E_{\vec m,\vec n}=A_{\vec m,\vec n} A_{\vec m,\vec n}^\dagger,
\ee
where
\be
\label{Adef}
A_{\vec m,\vec n}=F(\frac{X_1}{\Delta}+m_1)^{1/2} ... F(\frac{X_d}{\Delta}+m_d)^{1/2} F(\frac{Y_1}{\Delta}+n_1)^{1/2} ... F(\frac{Y_d}{\Delta}+n_d)^{1/2}.
\ee
We can verify that these form a POVM (namely, that the sum $\sum_{\vec m,\vec n} E_{\vec m,\vec n}$ is equal to the identity) by first summing over $n_d$ and using Eq.~(\ref{sumrule}) to show that $\sum_{n_d} F(\frac{Y_d}{\Delta}+n_d)=1$, then next summing over $n_{D-1}$ and so on.  Note also that the operators $E_{\vec m,\vec n}$ are all positive semi-definite as required for a POVM.

Let the $U_i$ act on some Hilbert space ${\cal H}$.
Given that this is a POVM, we can define some larger Hilbert space ${\cal H}'$ which is a direct sum of the original Hilbert space ${\cal H}$ with some auxiliary Hilbert space such that the following holds.  
\begin{lemma}
\label{fundprop}
 Let $\Pi$ be the projector acting on this larger space which projects onto the subspace corresponding to ${\cal H}$; then there are projectors $Q_{\vec m,\vec n}$ on this larger space such that $\Pi Q_{\vec m,\vec n} \Pi$, restricted to ${\cal H}$, equals $E_{\vec m,\vec n}$.
\begin{proof}
This statement is simply a finite-dimensional statement of Naimark's dilation theorem\cite{Naimark}.
\end{proof}
\end{lemma}

Define operators
\be
X'_i=\sum_{\vec m,\vec n} m_i\Delta Q_{\vec m,\vec n},
\ee
and
\be
Y'_i=\sum_{\vec m,\vec n} n_i \Delta Q_{\vec m,\vec n}.
\ee
These form a set of exactly commuting self-adjoint operators: $[X'_i,X'_j]=[X'_i,Y'_j]=[Y'_i,Y'_j]=0$.
Define
\be
V_i=X'_i+iY'_i,
\ee
so that the $V_i$ are exactly commuting and are normal.  To construct a set of exactly commuting unitaries, $U'_i$ from the $V'_i$, find a complex scalar $z$ with $|z|= 1$ such that $V_i+x z$ is invertible for all $i$ and all real $x$ with $0 < x \leq 1$; since the matrices are finite-dimensional, such a $z$ exists.  Then,
\be
\label{UpiDef}
U'_i=\lim_{x \rightarrow 0^+}{\rm polar}(V_i+x z).
\ee
Remark: in what follows, it is not that important that we find such a $z$ and take a limit as $x\rightarrow 0^+$; we could instead find any small $z$ such that $V_i+z$ is invertible for all $i$ and set $U'_i={\rm polar}(V_i+z)$.  We would then have to carry some additional error bounds dealing with the magnitude of the given $z$, but the proof would not change in any substantial way.  Alternately, we could simply arbitrarily map the $0$ eigenvalue to the point $1$ on the unit circle.

\begin{definition}
Define the map $\fm$ to map an $\epsilon$-sort torus ${\cal U}$ to an $\epsilon'$-local projector with projector $\Pi$ and unitaries $U'_1,...,U'_d$ following the procedure above, with $U'_i$ defined by Eq.~(\ref{UpiDef}), and with
\be
\label{deltadef}
\Delta=\sqrt{d \epsilon}.
\ee
\end{definition}

We will show in the next subsection in lemma (~\ref{epsplemma}) that the result is indeed an $\epsilon'$-local projector with
\be
\label{epsp}
\epsilon'=O(d \epsilon)^{1/4}.
\ee

\subsection{$\epsilon'$-Local Projector and Composition of Maps $\gm \circ \fm$}
In this section, we prove lemma (\ref{epsplemma}) showing that the result of the above definition is indeed an $\epsilon'$-local projector, and we also show that the composition of
maps $\gm \circ \fm$ is close to the identity in lemma (\ref{gmfm}).  Both these lemmas rely on similar preliminary results which we now give.

\begin{lemma}
\label{XiYi}
The operator
$\Pi X'_i \Pi$, restricted to the space ${\cal H}$, is equal to $X_i$, up to an error of $O(\Delta+d \epsilon \Delta^{-1})$ in operator norm for all $i$.
Similarly, $\Pi Y'_i \Pi$, again restricted to the space ${\cal H}$, equals $Y_i$, up to the same error of $O(\Delta+d \epsilon \Delta^{-1})$.

For the particular choice of $\Delta$ above, we have $O(\Delta+d \epsilon \Delta^{-1})=O(\sqrt{d \epsilon})$.
\begin{proof}
The operator $\Pi X'_i \Pi$, restricted to ${\cal H}$, is equal to
\be
\sum_{\vec m,\vec n} m_i \Delta E_{\vec m,\vec n}.
\ee
So, we wish to bound
\be
\label{diff1}
\Vert \sum_{\vec m,\vec n} m_i \Delta E_{\vec m,\vec n} - X_i \Vert.
\ee
Consider first the case $i=1$.  Then, 
\be
\sum_{\vec m,\vec n} m_1 \Delta E_{\vec m,\vec n}=\sum_{m_1} m_1 \Delta F(\frac{X_1}{\Delta}+m_1).
\ee
To show this, use $\sum_n F(\frac{X_j}{\Delta}+n)=\sum_n F(\frac{Y_j}{\Delta}+n)=1$, and first sum over $n_d$, then over $n_{d-1}$ and so on, until only the sum over $m_1$ is left.
Thus the difference (\ref{diff1}) equals
\be
\Vert \sum_{m_1} m_1 \Delta F(\frac{X_1}{\Delta}+m_1) - X_1 \Vert
\ee
which by definition of $F$ is bounded by $O(\Delta)$.

Now consider $i \neq 1$. 
We need to compute $\sum_{\vec m,\vec n} m_i \Delta E_{\vec m,\vec n}$.  Summing over $\vec n$ and over $m_{i+1},...,n_d$, this equals
\be
\label{miInt}
\sum_{m_1,...,m_i} m_i \Bigl( F(\frac{X_1}{\Delta}+m_1)^{1/2} ... F(\frac{X_i}{\Delta}+m_i)^{1/2}\Bigr)
\Bigl( F(\frac{X_i}{\Delta}+m_i)^{1/2} ... F(\frac{X_1}{\Delta}+m_1)^{1/2} \Bigr).
\ee
Summing over $m_i$ and using $\sum_{m_i} m_i \Delta F(\frac{X_i}{\Delta}+m_i) = X_i + O(\Delta)$, Eq.~(\ref{miInt}) equals
\be
\label{miInt2}
\sum_{m_1,...,m_{i-1}} m_i \Bigl( F(\frac{X_1}{\Delta}+m_1)^{1/2} ... F(\frac{X_{i-1}}{\Delta}+m_{i-1})^{1/2}\Bigr) (X_i+O(\Delta))
\Bigl( F(\frac{X_{i-1}}{\Delta}+m_{i-1})^{1/2} ... F(\frac{X_1}{\Delta}+m_1)^{1/2} \Bigr).
\ee

We first claim that
$\sum_{m_1,...,m_{i-1}} m_i \Bigl( F(\frac{X_1}{\Delta}+m_1)^{1/2} ... F(\frac{X_{i-1}}{\Delta}+m_{i-1})^{1/2}\Bigr) (O(\Delta))
\Bigl( F(\frac{X_{i-1}}{\Delta}+m_{i-1})^{1/2} ... F(\frac{X_1}{\Delta}+m_1)^{1/2} \Bigr)$ is itself bounded by $O(\Delta)$ in operator norm.  To see this, we  bound the trace of this operator with any matrix $\rho$ with trace norm bounded by $1$.  However, this is equal to the trace of
\be
\label{cpmapofrho}
\sum_{m_1,...,m_{i-1}} \Bigl( F(\frac{X_{i-1}}{\Delta}+m_{i-1})^{1/2} ... F(\frac{X_1}{\Delta}+m_1)^{1/2} \Bigr) \rho \Bigl( F(\frac{X_1}{\Delta}+m_1)^{1/2} ... F(\frac{X_{i-1}}{\Delta}+m_{i-1})^{1/2}\Bigr)
\ee
with some operator that is $O(\Delta)$.  Since the operator in Eq.~(\ref{cpmapofrho}) is given by a quantum channel (a completely-positive trace-preserving map) applied to $\rho$, the resulting operator still has trace norm bounded by $1$.  So, its trace with any operator with operator norm bounded by $O(\Delta)$ is bounded by $O(\Delta)$.  So, Eq.~(\ref{miInt2}) is $O(\Delta)$ plus
\be
\sum_{m_1,...,m_{i-1}} m_i \Bigl( F(\frac{X_1}{\Delta}+m_1)^{1/2} ... F(\frac{X_{i-1}}{\Delta}+m_{i-1})^{1/2}\Bigr) X_i
\Bigl( F(\frac{X_{i-1}}{\Delta}+m_{i-1})^{1/2} ... F(\frac{X_1}{\Delta}+m_1)^{1/2} \Bigr).
\ee

We next bound $\Vert [F(\frac{X_j}{\Delta}+m_i)^{1/2}, X_i \Vert$, using $\Vert [X_i,X_j] \Vert \leq \epsilon$.  
Let $g(t)$ be the Fourier transform of the function $F(...)^{1/2}$.  Since $F(...)^{1/2}$ is infinitely differentiable, $g(t)$ decays superpolynomially in $t$.
We have
\be
F(\frac{X_j}{\Delta}+m_j)^{1/2}=\int {\rm d}t g(t) \exp(i t(\frac{X_j}{\Delta} + m_i)),
\ee
and
\begin{eqnarray}
\Vert [F(\frac{X_j}{\Delta}+m_j)^{1/2}, X_i] \Vert & \leq & \int {\rm d}t |g(t)| 
\Vert [\exp(i t (\frac{X_j}{\Delta} + m_j)), X_i] \Vert \\ \nonumber
& \leq & 
\epsilon
\int {\rm d}t |g(t)|  |t| \Vert [\frac{X_j}{\Delta},X_i] \Vert
\\ \nonumber
&=& O(\epsilon \Delta^{-1}).
\end{eqnarray}
We used the decay of $g(t)$ to show that the above integral over $t$ converges.  In fact, we do not need superpolynomial decay, and a sufficiently fast polynomial would have sufficed.

Using this bound on the commutator, we can commute $X_i$ through $F(\frac{X_j}{\Delta}+m_j)^{1/2}$ for $j<i$ to obtain
\begin{eqnarray}
\sum_{m_1,...,m_{i-1}} m_i \Bigl( F(\frac{X_1}{\Delta}+m_1)^{1/2} ... F(\frac{X_{i-1}}{\Delta}+m_{i-1})^{1/2}\Bigr) X_i
\Bigl( F(\frac{X_{i-1}}{\Delta}+m_{i-1})^{1/2} ... F(\frac{X_1}{\Delta}+m_1)^{1/2} \Bigr)=X_i
+O(d \epsilon \Delta^{-1}).
\end{eqnarray}

So, Eq.~(\ref{miInt}) equals $X_i+O(\Delta+d\epsilon \Delta^{-1})$.

A similar proof can be used to bound the error $\Vert Y'_i-Y_i \Vert$.  In this case, we need to commute
$F(\frac{Y_i}{\Delta}+m_i)^{1/2}$ through $F(\frac{Y_j}{\Delta}+m_j)^{1/2}$ for $j<i$ and also commute
$F(\frac{Y_i}{\Delta}+m_i)^{1/2}$ through $F(\frac{X_j}{\Delta}+m_j)^{1/2}$ for all $j$.
\end{proof}
\end{lemma}
Remark: the above lemma can be thought of physically as follows: the POVM corresponds to a sequence of ``soft measurements", first of $X_1$, then of $X_2$, then $X_3$, and so on.  Each measurement gives some information on the value of the given $X_i$, and since the $X_i$ almost commute and the measurements are soft, they only weakly disturb subsequent measurements.

We need a related lemma:
\begin{lemma}
\label{XiYi2}
The operator
$\sum_{\vec m,\vec n} m^2_i \Delta^2 P_{\vec m,\vec n}$, is equal to $X^2_i$, up to an error of $O(\Delta+d \epsilon \Delta^{-1})$ in operator norm for all $i$.
Similarly, $\sum_{\vec m,\vec n} n^2_i \Delta P_{\vec m,\vec n}$, again restricted to the space ${\cal H}$, equals $Y^2_i$, up to the same error of $O(\Delta+d \epsilon \Delta^{-1})$.
\end{lemma}
The proof of this lemma is almost the same as that of lemma (\ref{XiYi}), so we do not repeat it.

Next we show that 
\begin{lemma}
\label{ugly}
\be
\Vert \Pi ( U'_i - V_i ) \Pi \Vert
\leq O(\Delta+d \epsilon \Delta^{-1})^{1/2}.
\ee

For the given $\Delta$, this is $O(d \epsilon)^{1/4}$.
\begin{proof}
Let $Q$ project onto the eigenspace of $V_i$ with eigenvalues whose absolute value squared is outside the interval $[1-x,1+x]$ for some real number $x$.
Then,
\begin{eqnarray}
\Pi Q \Pi & = & \sum_{\vec m,\vec n; (m_i^2 + n_i^2) \Delta^2 \not \in [1-x,1+x]} E_{\vec m,\vec n}.
\end{eqnarray}

Using lemma (\ref{XiYi2}), and $X_i^2+Y_i^2=I$, we have
$\sum_{\vec m,\vec n}  (m_i^2 + n_i^2) \Delta^2 E_{\vec m,\vec n}=I+ \leq O(\Delta+d \epsilon \Delta^{-1})$, and so
\be
\Vert \sum_{\vec m,\vec n}  \Bigl((m_i^2 + n_i^2) \Delta^2-1\Bigr) E_{\vec m,\vec n} \Vert \leq O(\Delta+d \epsilon \Delta^{-1}).
\ee
The operator in the left-hand side of the above equation is lower bounded by $x \Pi Q \Pi$ restricted to ${\cal H}$.  So,
\be
\Vert \Pi Q \Pi \Vert
\leq \frac{1}{x}O(\Delta+d \epsilon \Delta^{-1}).
\ee

We have $\Vert (1-Q) (U'_i-V_i ) (1-Q) \Vert \leq x$.  So,
$\Vert \Pi ( U'_i - V_i ) \Pi \Vert \leq \Vert \Pi (1-Q) (U'_i-V_i ) (1-Q) \Pi \Vert + \Vert \Pi Q (U'_i -V_i) Q \Pi \Vert \leq
x+{\rm const.} \times \Vert \Pi Q \Pi \Vert$.
Hence,
\be
\Vert \Pi ( U'_i - V_i ) \Pi \Vert  \leq x +  \frac{1}{x}O(\Delta+d \epsilon \Delta^{-1}).
\ee
Picking $x=\sqrt{\Delta+d \epsilon \Delta^{-1}}$, we get that
\be
\Vert \Pi ( U'_i - V_i ) \Pi \Vert
\leq O(\Delta+d \epsilon \Delta^{-1})^{1/2}.
\ee
\end{proof}
\end{lemma}


We can now prove that
\begin{lemma}
\label{epsplemma}
For all $i$,
\be
\Vert \Pi,U'_i \Vert \leq O(d \epsilon)^{1/4}.
\ee
\begin{proof}
By lemma (\ref{ugly}) and a triangle inequality, 
\begin{eqnarray}
\Vert \Pi,U'_i \Vert & \leq & \Vert [\Pi,V_i] \Vert+
 O(d\epsilon)^{1/4} \\ \nonumber
& \leq & \Vert [\Pi,X'_i] \Vert + \Vert [\Pi,Y'_i] \Vert + O(d \epsilon)^{1/4}.
\end{eqnarray}

We now bound $\Vert [\Pi,X'_i] \Vert$.  Let
$(X'_i)^{11}=\Pi X'_i \Pi$, let $(X'_i)^{12}=\Pi X'_i (1-\Pi)$, and let $(X'_i)^{21}=(1-\Pi) X'_i \Pi$.
Then, to bound $\Vert [\Pi,X'_i] \Vert$, it suffices to bound $$\Vert (X'_i)^{12} \Vert = \sqrt{\Vert (X'_i)^{12} (X'_i)^{21} \Vert}=\Vert \Pi (X'_i)^2 \Pi - (\Pi X'_i \Pi)^2 \Vert.$$
However, by lemmas (\ref{XiYi},\ref{XiYi2}), $\Pi X_i \Pi=X_i+O(d \epsilon)^{1/2}$ and $\Pi (X'_i)^2 \Pi=X_i^2+O(d \epsilon)^{1/2}$ so
$\Vert \Pi (X'_i)^2 \Pi - (\Pi X'_i \Pi)^2 \Vert=O(d \epsilon)^{1/2}$ and so
 $\Vert [\Pi,X'_i] \Vert=O(d \epsilon)^1/4)$.

A similar bound for $Y'_i$ holds with a similar proof.
\end{proof}
\end{lemma}

Finally, as a corollary of lemma (\ref{ugly}) and lemma (\ref{mapsto}), note that
\begin{lemma}
\label{gmfm}
Given any $\epsilon$-soft torus ${\cal U}$ with unitaries $U_1,...,U_d$, the composition of maps $\gm \circ \fm$ applied to ${\cal U}$ gives the same unitaries up to an error
that is
$O(d \epsilon)^{1/4}$ in operator norm.
The resulting $\fm \circ \gm$ is an $O(d^{1/2} \epsilon^{1/2})$ soft torus.
\end{lemma}

\subsection{Composition of Maps $\fm \circ \gm$}
We now show that the composition of maps $\fm \circ \gm$ is close to the identity in lemma \ref{fmgm}.  Note that if ${\cal P}$ is an $\epsilon$-local projector, then
$\fm \circ \gm {\cal P}$ will be a $\delta$-local projector for $\delta=O(d^{1/4} \epsilon^{1/2})$, and $\delta$ may be bigger than $\epsilon$ for small $\epsilon$.  So, when we say that the composition of maps is close to the identity, we measure the distance treating these as $\delta$-local projectors.

The distance between two local projectors has the following useful property:
\begin{lemma}
\label{swindle}
Consider an $\epsilon$-local projector ${\cal P}$ corresponding to unitaries $U_1,...,U_d$ and projector $P$ and an $\epsilon$-local projector ${\cal Q}$ corresponding to unitaries $V_1,...,V_d$,
such that $P$ and $Q$ have the same rank.  Then
\be
{\rm dist}({\cal P},{\cal Q})\leq {\rm dist}(\gm {\cal P}, \gm {\cal Q})+O(\epsilon).
\ee
\begin{proof}
Refer to definition \ref{LocHamDistDef}.
We will define a pair of trivial local projectors, called ${\cal R}$ and ${\cal S}$, as in that definition.
We fix the unitaries in ${\cal R}$ to be the unitaries $V_1,...,V_d$ and we fix the unitaries in ${\cal S}$ to be the unitaries $U_1,...,U_d$.  This then fulfills the condition
 $*$ in that definition.  Let the projectors $R$ and $S$ in ${\cal R}$ and ${\cal S}$ both be equal to zero.

Let the tuple of unitaries corresponding to ${\cal P}+{\cal R}$ be $u_1,...,u_d$ and let  tuple of unitaries corresponding to ${\cal Q}+{\cal S}$ be $v_1,...,v_d$.
Let $p,q$ be the projectors in ${\cal P}+{\cal R}$ and ${\cal Q}+{\cal S}$, respectively.
We compute $d({\cal P}+{\cal R},{\cal Q}+{\cal S})$, which was defined to be the minimum over all unitaries $Y$ such that $Y^\dagger p Y = q$ of
\be
{\rm max}_i \Vert Y^\dagger u_i Y - v_i \Vert.
\ee
Let us make a basis transformation so that we write
\be
u_i=\begin{pmatrix} u_i^{11} & u_i^{12} \\ u_i^{21} & u_i^{22} \end{pmatrix},
\ee
where here we write $u_i$ as a block matrix, with
\be
p=\begin{pmatrix} I & 0 \\ 0 & 0 \end{pmatrix}.
\ee
Since the distance is invariant under simultaneous conjugation of $v_i$ and $q$ by any unitary, let us without loss of generality assume that $p=q$.  Then, we will write $v_i$ in the same basis as
\be
v_i=\begin{pmatrix} v_i^{11} & v_i^{12} \\ v_i^{21} & v_i^{22} \end{pmatrix},
\ee

We have 
\be
\Vert u_i- \begin{pmatrix} u_i^{11} & 0 \\ 0 & u_i^{22} \end{pmatrix} \Vert \leq O(\epsilon),
\ee
and a similar bound for $v_i$.  Let $d_1$ be the minimum over matrices $Y_1$ of
\be
{\rm max}_i \Vert Y_0^\dagger u_i^{11} Y_0 - v_i^{11} \Vert,
\ee
and let $d_2$ be the minimum over matrices $Y_2$ of
\be
{\rm max}_i \Vert Y_1^\dagger u_i^{22} Y_1 - v_i^{22} \Vert.
\ee
Then, we can upper bound $d({\cal P}+{\cal R},{\cal Q}+{\cal S})\leq {\rm max}(d_1,d_2)+O(\epsilon)$ by choosing $Y$ to be the direct sum of $Y_1$ and $Y_2$ and minimizing over $Y_1,Y_2$ separately.
However,  $\Vert u_i^{11}-{\rm polar}(u_i^{11}) \Vert \leq O(\epsilon^2)$ by Eq.~(\ref{closetopolar}), and similarly for $v_i^{11}$, so
$d_1={\rm dist}(\gm {\cal P},\gm {\cal Q})+O(\epsilon^2)$.

Let us write
\be
U_i=\begin{pmatrix} U_i^{11} & U_i^{12} \\ U_i^{21} & U_i^{22} \end{pmatrix},
\ee
in a basis where the projector in ${\cal P}$ is equal to
\be
\begin{pmatrix} I & 0 \\ 0 & 0 \end{pmatrix},
\ee
and also write $V_i$ similarly.  Note that $U_i^{12}$ and $U_i^{21}$ are both $O(\epsilon)$.  Then
\be
u_i^{22}=\begin{pmatrix} {\rm polar}(U_i^{22}) &0 & 0 \\ 0 & {\rm polar}(V_i^{11}) & 0 \\ 0 & 0 & {\rm polar}(V_i^{22}) \end{pmatrix} + O(\epsilon),
\ee
and
\be
v_i^{22}=\begin{pmatrix} {\rm polar}(V_i^{22}) &0 & 0 \\ 0 & {\rm polar}(U_i^{11}) & 0 \\ 0 & 0 & {\rm polar}(U_i^{22}) \end{pmatrix} + O(\epsilon).
\ee
To upper bound $d_2$, consider the matrix $Y_2$ that is
\be
Y_1=\begin{pmatrix} 0 & 0 & I \\ 0 & I & 0 \\ I & 0 & 0 \end{pmatrix}.
\ee
For this $Y_2$, we have
${\rm max}_i \Vert Y_1^\dagger u_i^{22} Y_1 - v_i^{22} \Vert={\rm dist}(\gm {\cal P},\gm {\cal Q})+O(\epsilon)$.

From these bounds on $d_1,d_2$, the claim follows.
\end{proof}
\end{lemma}

\begin{lemma}
\label{fmgm}
Let ${\cal P}$ be an $\epsilon$-local projector.
For sufficiently small $\epsilon$,
we have
\be
\label{bf}
{\rm dist}({\cal P},\fm \circ \gm {\cal P})=O(d \epsilon)^{1/4},
\ee
where we $\fm \circ \gm$ is an $O(d^{1/4} \epsilon^{1/2}$-local projector.
\begin{proof}
By lemma (\ref{mapsto}), the map $\gm$ maps ${\cal P}$ to an $O(\epsilon^2)$-soft torus for sufficiently small $\epsilon$.
By lemma (\ref{ugly}), the map $\fm$ maps this soft torus to an $O(d^{1/4} \epsilon^{1/2})$-local projector ${\cal Q}$, such that
${\rm dist}(\gm {\cal P},\gm {\cal Q})=O(d^{1/4} \epsilon^{1/2})$.
Hence
For the given $\Delta$, this is $O(d \epsilon)^{1/4}$.
Hence, by lemma \ref{swindle}, Eq.~(\ref{bf}) follows.
\end{proof}
\end{lemma}

\section{Map Between $\epsilon$-local Projectors and Vector Bundles}
\label{maps2}
Rather than considering vector bundles directly, we consider functions $E$ from the $d$-dimensional torus to the space of projectors in some finite-dimensional complex vector space.  We write parametrize the torus by angles $\theta_1,...,\theta_d$.  Then, $E(\vec \theta)$ is a projector which depends periodically upon the angles $\theta_i$ with period $2\pi$.

We define a distance ${\rm dist}(\vec \theta,\vec \theta')$ to be ${\rm max}({\rm dist}(\theta_i,\theta'_i))$, where
given any two angles $\theta,\theta'$, we define ${\rm dist}(\theta,\theta')=({\rm min}_n(|\theta_i-\theta'_i+2 \pm n|))$, where the min is taken over all integers $n$.

Throughout, we consider functions $E$ which are infinitely differentiable.  Since the domain is compact, the first derivative is bounded and so we have
\be
\label{Lipschitz}
\frac{\Vert E(\vec \theta)-E(\vec \theta') \Vert}{{\rm dist}(\vec \theta,\vec \theta')} \leq K
\ee
for some constant $K$.  The particular value of this constant $K$ will play a role in some of the estimates later.

In this section we define maps $\um,\vm$ from such projector-valued functions to local projectors and vice-versa.  We then show that the composition of maps $\um \circ \vm$ or $\vm \circ \um$ is close to the identity.  Whenever we refer to a projector-valued function, we mean a projector-valued function depending periodically upon angles $\theta_1,...,\theta_d$.

\subsection{Definition of Map $\um$}
\begin{definition}
Define a map $\um$ from such
projector-valued functions $E$ to 
$\epsilon$-local projectors as follows.  Let $E(\vec \theta)$ be a projector acting on a $D$-dimensional space, with basis vectors $v_1,...,v_D$.  Let $N$ be an arbitrary integer.
Let $\um P$ be the $\epsilon$-local projector defined as follows.
The matrices act on a space of dimension $D N^d$.
This space has basis vectors which we write as
\be
\label{basisdef1}
|\vec n; a \rangle,
\ee
where
$a=1,...,D$ and the vector $\vec n$ refers to $n_1,...,n_d$ with $0 \leq n_i \leq N-1$.
This notation indicates a tensor product decomposition of the space into an $N^d$-dimensional space and a  $D$-dimensional space:
\be
| \vec n; a \rangle=|\vec n \rangle \otimes v_a.
\ee
This tensor product decomposition is used in Eq.~(\ref{Pdef}).

Let $\vec x(i)$ denote the vector with a $1$ in the $i$-th position and a zero everywhere else.
We define the matrices $U_i$ by
\be
\label{UiDef1}
U_i=\sum_{\vec n;a} |\vec n+\vec x(i);a\rangle \langle \vec n; a |=\sum_{\vec n} |\vec n+\vec x(i)\rangle \langle \vec n| \otimes I.
\ee
The addition $\vec n+\vec x(i)$ is mod $N$, giving the tuple $n_1,...,n_{i-1},n_i+1 \, {\rm mod} \, N,n_{i+1},...,n_d$.
That is, $U_i$ increases $n_i$ by $1$ (mod $N$) and acts as the identity on the rest of the space.

We define the projector for the $\epsilon$-local projector to be $\hat P$
\be
\label{Pdef}
\hat P=\sum_{\vec n} |\vec n\rangle \langle \vec n| \otimes E(2 \pi \frac{\vec n}{N}),
\ee
where $2\pi \frac{\vec n}{N}$ refers to 
$2\pi \frac{n_1}{N},...,2\pi \frac{n_d}{N})$.
\end{definition}

\begin{lemma}
Assume $E$ obeys Eq.~(\ref{Lipschitz}) for given $K$.
For given $N$, the map $\um$ above maps such projectors to $\epsilon$-local projectors
with
\be
\epsilon \leq 2 \pi \frac{K}{N}.
\ee
\begin{proof}
We bound $\Vert [ U_i,P] \Vert$.  This is equal to the maximum over $\vec n$ of
\be
\Vert E(2\pi \frac{\vec n}{N})-E(2\pi \frac{\vec n+\vec x(i)}{N}) \Vert.
\ee
By Eq.~(\ref{Lipschitz}), this is bounded by $2\pi K/N$.
\end{proof}
\end{lemma}

\subsection{Definition of Map $\vm$}
Given an $\epsilon$-local projector with commuting unitaries $U_1,...,U_d$ we will work in a basis that diagonalize all of the $U_i$ simultaneously,
so that every basis element $v$ obeys $U_i v = z_i v$ for some $|z_i|=1$.
In what follows we choose some real number $R>0$.
From the projector $P$, we now define a new operator$H_{loc}$ with the property
that $\Vert P - H_{loc} \Vert$ is small, going to zero as $\epsilon/R$ goes to zero, and such that
if $v_1,v_2$ are basis elements with $U_i v_1=z^1_i v_1$ and $U_i v_2=z^2_i v_2$ and such that
$z_1=\exp(i \phi_1)$ and $z_2=\exp(i \phi_2)$ with ${\rm dist}(\phi_1,\phi_2) \geq R$,
then
\be
\label{cons}
\langle v_1 |  H_{loc} v_2\rangle=0.
\ee

The intuition for this property of $H_{loc}$ is that we think of the basis elements as corresponding to ``sites" of some tight-binding model in a $d$-dimensional torus and $P$ is some Hamiltonian for free electrons hopping between these sites. We replace $P$ by some approximation $H_{loc}$ such that
$H_{loc}$ is now ``short-range": it has strictly zero matrix elements between sufficiently far separated sites.
The particular choice of $R$ above is not too important so long as $R$ is sufficiently small; the important thing is that the matrix element vanishes if $z_1$ is close to $-z_2$.

Note that $H_{loc}$ need not be a projector.  However, will show later that for sufficiently small $\epsilon$, the eigenvalues of $H_{loc}$ are bounded away from $1/2$.  We refer to $H_{loc}$ as an $R$-strictly local Hamiltonian.

To define $H_{loc}$, first we define
\be
X_i=\frac{U_i +  U_i^\dagger}{2},
\ee
\be
Y_i=\frac{U_i-U_i^\dagger}{2}.
\ee
Note that $[X_i,X_j]=[X_i,Y_j]=[Y_i,Y_j]=0$.

To construct $H_{loc}$, we will instead construct an operator such that, for some scalar $S$, if
$|{\rm Re}(z_1)-{\rm Re}(z_2)| \geq S$ or
$|{\rm Im}(z_1)-{\rm Im}(z_2)| \geq S$, then
\be
\label{consS}
\langle v_1| H_{loc} v_2\rangle=0.
\ee
By choosing $S$ sufficiently small, this assumption will imply that Eq.~(\ref{cons}) holds if $z_1=\exp(i \phi_1)$ and $z_2=\exp(i \phi_2)$ with ${\rm dist}(\phi_1,\phi_2) \geq R$.
A sufficiently small $S$ can be chosen by picking $S$ equal to a constant times $R$.

Then,
define
\be
H_{loc} = S^{2d} \int {\rm d}a_1 {\rm d}b_1 {\rm d}a_2 {\rm d}b_2 ... {\rm d}a_d {\rm d}b_d
\exp(i \sum_j (a_j X_j + b_j Y_j)) P \exp(-i \sum_j (a_j X_j+b_j Y_j)) \prod_j (F(S a_j) F(S b_j)),
\ee
where the function $F$ is defined to be the Fourier transform of a function $\tilde F$, where $\tilde F$ can be any given function with the property that $\tilde F$
is sufficiently smooth that $F(t)$ is bounded by a constant divided by $t^3$ and such that $\tilde F(0)=1$ and $\tilde F(\omega)$ vanishes for $|\omega| \geq 1$.
This last property implies that Eq.~(\ref{cons}) will hold for the given $v_1,v_2$.
Then,
we can bound
\begin{lemma}
\label{closetoPlemma}
\be
\label{closetoP}
\Vert H_{loc}-P \Vert \leq O(\epsilon).
\ee

$H_{loc}$ is $O(\epsilon)$-local with respect to the $U_i$.
\begin{proof}
Since $\tilde F(0)=1$, we have
\be
P= S^{2d} \int (\prod_i {\rm d}a_i {\rm d}b_i) 
P \prod_j (F(Sa_j) F(Sb_j)),
\ee
so by a triangle inequality
\begin{eqnarray}
\Vert H_{loc}-P \Vert &\leq  &S^{2d} \int (\prod_i {\rm d}a_i {\rm d}b_i) 
\Vert\exp(i \sum_j (a_j X_j + b_j Y_j)) P \exp(-i \sum_j (a_j X_j+b_j Y_j))- P \Vert \prod_j (F(Sa_j) F(Sb_j)) \nonumber \\
&\leq & S^{2d} \epsilon \int (\prod_i {\rm d}a_i {\rm d}b_i) 
\sum_j |a_j|+|b_j|) \prod_j (F(Sa_j) F(Sb_j)).
\end{eqnarray}
Because of the decay of $F$, the above integral converges and is bounded by a constant times $\epsilon/S$.

The fact that $\Vert [H_{loc},U_i] \Vert$ is bounded by $O(\epsilon)$ follows from the bound on $\Vert H_{loc}-P \Vert$ and from the fact that $P$ is $\epsilon$-local.
\end{proof}
\end{lemma}

Having defined $H_{loc}$, we next define an operator $H_{loc}(\vec \theta)$, which depends periodically upon angles $\theta_1,...,\theta_d$.
Physically, this definition corresponds to a twist in boundary conditions by angles $\theta_1,...,\theta_d$.  The torus $\vec \theta$ is sometimes called the ``flux torus"\cite{fluxtorus}.
Using the definition of $H_{loc}(\vec \theta)$, we define a projector-valued function $E$.

Let us first give a general definition of ``twisted boundary conditions" for an arbitrary operator, before specifying to $H_{loc}$.
Without loss of generality we suppose that none of the matrices $U_i$ have $+1$ in their spectrum (since they are finite-dimensional matrices, we can always achieve this by replacing $U_i$ by $U_i$ multiplied by some phase).  We remark that this supposition is just a technical detail to help simplify the definition of twisted boundary conditions; if we had not made this assumption, we would have to also define below what to do if $z_i=1$ or$z_j=1$.
\begin{definition}
\label{tOdefine}
Let $O$ be any that is $\pi/2$-strictly local.
operator such that the following property holds:

Then, define the operator $O(\vec \theta)$ by its matrix elements.
Let $v_1,v_2$ be eigenvevectors of all the $U_i$, with $U_i v_1=z^1_i v_1$ and $U_i v_2=z^2_i v_2$, and $z^1_i=-z^2_i$.
For each $i=1,...,d$, consider the shortest path on the unit circle from $z^2_i$ to $z^1_i$.  If this path contains the point $+1$ on the unit circle and moves in a counter-clockwise direction, then set $\omega_i=\exp(i \theta_i)$; if it contains $+1$ and moves in a clockwise direction then set $\omega_i=\exp(-i \theta_i)$.  Otherwise set $\omega_i=1$.  Then, let
$\langle v_1 |O(\vec \theta) v_2\rangle=(\prod_i \omega_i)  \langle v_1| O v_2 \rangle$.

We refer to $O(\vec \theta)$ as ``$O$ with twisted boundary conditions".
\end{definition}
Note that due to the condition of $\pi/2$-strict locality, the choice of shortest path is unambiguous.  In fact, $\pi$-strict locality would have sufficed, but for use later we pick $\pi/2$.

We now define $H_{loc}(\vec \theta)$.

\begin{definition}
\label{twistdefine}
Let $R \leq \pi/2$ and construct $H_{loc}$.
Given $H_{loc}$, define $H_{loc}(\vec \theta)$ as above.
Define $E(\vec \theta)$ to project onto the eigenspace of $H_{loc}(\vec \theta)$ with eigenvalue more than $1/2$.

We define a map $\vm$ from $\epsilon$-local projectors to projector-valued functions by setting $\vm {\cal P}$ to be the projector-valued function $E$ obtained by first constructing the operator $H_{loc}$ above and then constructing $H_{loc}(\theta)$ and $E(\vec \theta)$ as in this definition.
\end{definition}
The above definition constructs a projector onto an eigenspace of a particular operator $H_{loc}(\vec \theta)$ with eigenvalue more than $1/2$.  It is not a priori obvious that the dimension of this
eigenspace is independent of $\vec \theta$.  However, next we prove that for sufficiently small $\epsilon$ that the operator $H_{loc}(\vec \theta)$ is in fact close to a projector, and so we can bound its eigenvalues away form $1/2$.
We show after that in lemma \ref{PLip}  that the projector-valued function $E$ obeys the Lipschitz condition (\ref{Lipschitz}) and give bounds on $K$.


To prove that $H_{loc}(\vec \theta)$ is close to a projector, we need a technical lemma:
\begin{lemma}
For any $O$ fulling the conditions of definition (\ref{tOdefine}), we have
\be
\label{tNbound}
\Vert O(\vec \theta) \Vert \leq 2^d \Vert O \Vert.
\ee
\begin{proof}
We decompose $O$ as a sum of $2^d$ different operators $O_{m_1,...,m_d}$ with each $m_i$ being equal to either $0$ or $1$:
\be
O=\sum_{\{m_1,...,m_d\}} O_{m_1,...,m_d}.
\ee

Without loss of generality, let 
us suppose that none of the matrices $U_i$ have $-1$ in their spectrum.  This is in addition to our previous assumption that that do not have $+1$ in their spectrum, and again can be achieved by multiplying the matrices by a scalar and again is just a technical detail to simplify the proof.

The $O_{m_1,...,m_d}$ are defined by their matrix elements.
Let $v_1,v_2$ be eigenvectors of all the $U_i$, with $U_i v_1=z^1_i v_1$ and $U_i v_2=z^2_i v_2$, and $z^1_i=-z^2_i$.
Define integers $n_1,...,n_d$ as follows.
For each $i=1,...,d$, consider the shortest path on the unit circle from $z^2_i$ to $z^1_i$.  If this path contains the point $1$ on the unit circle or it contains the point $-1$ on the unit circle, then set $n_i=1$.  Otherwise set $n_i=0$.  Then, let
$\langle v_1 |O_{m_1,...,m_d} v_2\rangle=(\prod_i \omega_i)  \langle v_1| O v_2 \rangle$ if $m_i=n_i$ for all $i$ and otherwise site $\langle v_1 |O_{m_1,...,m_d} v_2\rangle=0$.

Then, for each $O_{m_1,...,m_d}$ define operators $O_{m_1,...,m_d}(\vec \theta)$ to be $O_{m_1,...,m_d}$ with twisted boundary conditions.
We claim that
\be
\Vert O_{m_1,...,m_d}(\vec \theta) \Vert = \Vert O_{m_1,...,m_d} \Vert.
\ee
This claim can be shown as follows.  We just show this for $d=1$ as $d>1$ can be handled similarly, just introducing more projectors.  Suppose $m_1=1$.  Let $\Pi_L$ project onto eigenvectors of $U_1$ with eigenvalues $z_1=\exp(i \phi_1)$ with $\phi_1 \in (\pi/2,3\pi/2)$ and let $\Pi_L$ project onto eigenvectors of $U_1$ with eigenvalues $z_1=\exp(i \phi_1)$ with $\phi_1 \in (-\pi/2,\pi/2)$. 
Let $\Pi_T$ project onto eigenvectors of $U_1$ with eigenvalues $z_1=\exp(i \phi_1)$ with $\phi_1 \in (0,\pi)$ and let $\Pi_L$ project onto eigenvectors of $U_1$ with eigenvalues $z_1=\exp(i \phi_1)$ with $\phi_1 \in (\pi,2\pi)$.
 Then, for $m_1=1$,
$O_{m_1}(\theta_1)=\Pi_L  O_{m_1} \Pi_L + \exp(i\Pi_T \theta_1)(\Pi_R O_{m_1}\Pi_R)\exp(-i\Pi_T \theta_1)$, and so $\Vert O_{m_1}(\theta)\leq {\rm max}(\Vert \Pi_L  O_{m_1} \Pi_L \Vert, \Vert \exp(i\Pi_T \theta_1)(\Pi_R O_{m_1}\Pi_R)\exp(-i\Pi_T \theta_1) \Vert) = {\rm max}(\Vert \Pi_L  O_{m_1} \Pi_L \Vert, \Vert \Pi_R O_{m_1}\Pi_R \Vert) \leq \Vert O_{m_1} \Vert$.
For $m_1=0$, $O_{m_1}(\theta_1)=O_{m_1}$.

In the above paragraph, a key role is played by the $\pi/2$-strict locality.  The reader should consider the decomposition of the space into the ranges of operators $\Pi_L,\Pi_R,\Pi_T,\Pi_B$ and see which spaces the operator $O_{m_1}$ can have matrix elements between.

Also,
\be
O(\vec \theta)=\sum_{\{m_1,...,m_d\}} O_{m_1,...,m_d}(\vec \theta).
\ee
So by a triangle inequality,
\be
\label{eqsum}
\Vert O(\vec \theta) \Vert \leq \sum_{\{m_1,...,m_d\}} \Vert O_{m_1,...,m_d} \Vert \Vert.
\ee

We claim that $\Vert O_{m_1,...,m_d} \Vert \leq \Vert O \Vert$.  Given this claim and Eq.~(\ref{eqsum}), Eq.~(\ref{tNbound}) follows.
To show this claim, consider first the case $d=1$ and $m_1=1$.  Then, as above, $O_{m_1}\leq {\rm max}(\Vert \Pi_L O_{m_1} \Pi_L \Vert,\Vert \Pi_R O_{m_1} \Pi_R \Vert)$.  Let us bound $\Vert \Pi_L O_{m_1} \Pi_L \Vert$, and the bound with $L$ replaced by $R$ will be similar.  This is equal to $\Vert \Pi_L \Pi_T P \Pi_B \Pi_L+\Pi_L \Pi_B  P \Pi_T \Pi_L \Vert$.  This norm is bounded by the max of the norm of the two terms in the sum, and each norm is bounded $\Vert O \Vert$.  Now consider $m_1=0$.  This is equal to $\Vert \Pi_T O \Pi_T + \Pi_B O \Pi_B \Vert \leq \Vert O \Vert$.
\end{proof}
\end{lemma}


We can now show that $H_{loc}(\theta)$ is close to a projector
\begin{lemma}
\label{HIsAppProj}
\be
\Vert (H_{loc}(\vec \theta))^2-H_{loc}(\vec \theta) \Vert \leq  2^d O(\epsilon).
\ee
\begin{proof}
We have chosen $R \leq \pi/4$.  So, $H_{loc}^2$ fulfills the assumptions of definition (\ref{tOdefine}) and in fact we find that
\be
(H_{loc}^2)(\vec \theta)=(H_{loc}(\vec \theta))^2.
\ee
That is, it does not matter whether we square the operator and then twist the boundary conditions or twist and then square.

Further, let $X=H_{loc}^2-H_{loc}$.  Then,
$H_{loc}(\theta))^2-H_{loc}(\theta)=X(\theta)$.
By lemma \ref{closetoPlemma}, $H_{loc}^2-H_{loc}=P^2-P+O(\epsilon)$ and since $P^2=P$, $\Vert X \Vert = O(\epsilon)$.
Hence, by lemma \ref{tNbound}, $\Vert X(\theta) \Vert \leq 2^d O(\epsilon)$.
\end{proof}
\end{lemma}

Also, we show that $H_{loc}(\vec \theta)$ almost commutes with all the $U_i$:
\begin{lemma}
\label{twistedAlmostCommutes}
\be
\Vert [H_{loc}(\vec \theta),U_i ] \Vert \leq \epsilon 2^d.
\ee
\begin{proof}
Let $C=[H_{loc},U_i]$.  Defined the operator $C(\vec \theta)$ with twisted boundary conditions as before.
We have $[H_{loc}(\vec \theta),U_i ]=C(\vec \theta)$.  By Eq.~(\ref{tNbound}), $\Vert C(\vec \theta) \Vert \leq 2^d \Vert C \Vert \leq 2^d \epsilon$.
\end{proof}
\end{lemma}

\begin{lemma}
\label{PLip}
For sufficiently small $\epsilon$, the rank of the projector $E(\vec \theta)$ is independent of $\vec \theta$, and $E$ is infinitely differentiable, and $E$ obeys Eq.~(\ref{Lipschitz}) with $K$ given by
\be
K=d\frac{1+2^d O(\epsilon)}{1-2^d O(\epsilon)}.
\ee
In fact, the projector obeys the stronger requirement that
\be
\Vert \partial_{\theta_i} E(\vec \theta) \Vert \leq \frac{1+2^d O(\epsilon)}{1-2^d O(\epsilon)}.
\ee
\begin{proof}
By lemma \ref{HIsAppProj}, for sufficiently small $\epsilon$, the eigenvalues of $H_{loc}(\vec \theta)$ are uniformly bounded away from $1/2$, so the rank of $E(\vec \theta)$ is independent of $\vec \theta$.  Given this uniform bound, and given that $H_{loc}(\vec \theta)$ is infinitely differentiable, so is $E$ for small enough $\epsilon$.

Given an operator $O$, which is strictly $\pi/2$-local, and some given $\vec \theta$, let $O'=O(\vec \theta)$.  Let $O''=O'(\vec \phi)$.  Then, $O''=O(\vec theta+\vec \phi)$.
So, for given $\vec \theta$, to compute $\partial_{\theta_i} H_{loc}(\vec \theta)$, let $O'=H_{loc}(\vec \theta)$, so that $\partial_{\theta_i} H_{loc}(\vec \theta)=\partial_{\phi_i} O'(\vec \phi)$, where the derivative on the left is taken at the given $\vec \theta$ and the derivative on the right is taken at $\vec \phi=0$.  However, at $\vec \phi=0$,
$\Vert \partial_{\phi_i} O'(\vec \phi) \Vert \leq \Vert O' \Vert$.  Hence
$\Vert \partial_{\theta_i} H_{loc}(\vec \theta) \Vert \leq \Vert H_{loc}(\vec \theta)$.  By lemma \ref{HIsAppProj}, $\Vert H_{loc}(\vec \theta) \Vert \leq 1+2^d O(\epsilon)$.  So,
\be
\Vert \partial_{\theta_i} H_{loc}(\vec \theta) \Vert \leq 1+2^d O(\epsilon).
\ee

The gap in the spectrum between the eigenvalues of $H_{loc}(\vec \theta)$ with eigenvalue less than $1/2$ and those with eigenvalue greater than $1/2$ is at least $1-2^d O(\epsilon)$.
So, using this gap and the bound on 
$\Vert \partial_{\theta_i} H_{loc}(\vec \theta) \Vert$, we get
$\Vert \partial_{\theta_i} E(\vec \theta) \Vert \leq (1+2^d O(\epsilon))/(1-2^d O(\epsilon))$.
\end{proof}
\end{lemma}

\section{The Composition of Maps $\um\circ \vm$}
We now compute the composition of maps $\um\circ \vm$, mapping an $\epsilon$-local projector ${\cal P}$ to a projector-valued function $E$, and back to a $2^d O(\epsilon)$-local projector.  The result will depend upon the parameter $N$ used to define the map $u$ and upon the parameter $R$.
We will pick $N$ to be $\lceil 1/\epsilon \rceil$ and we pick any fixed $R \leq \pi/2$.
Lemma \ref{umvm} bounds the distance of this composition of maps from the identity.

From the $\epsilon$-local projector ${\cal P}$, we obtain an $R$-strictly local Hamiltonian $H_{loc}$.  
Using the procedure above, we construct a projector $P$.

Let $|v_a\rangle$ for $a=1,...,D$ be a basis for the space that $P$ acts on.
Let us choose these basis vector $|v_a \rangle$ to be eigenvectors of the operators $U_i$.
The map $u$ produces unitaries whose eigenvalues are Fourier modes: for each tuple $\vec m$, there is a $D$-dimensional eigenspace of the $U_i$ with basis
\be
\label{newbasis}
| \vec m; a\rangle_F \equiv
\frac{1}{N^{D/2}}
\sum_{\vec n} \exp(i \vec \theta \cdot \vec m) | \vec n; a \rangle,
\ee
where the sum is over tuples $\vec n$ and where $\theta_i=2\pi \frac{n_i}{N}$.  The vectors $|\vec n; a \rangle$ are as defined in Eq.~(\ref{basisdef1}).
These basis vectors in Eq.~(\ref{newbasis}) are eigenvectors of $U_i$ with eigenvalues $\exp(-i 2\pi n_i/N)$.
The subscript $F$ is used to indicate that this is a different basis from (\ref{basisdef1}); in some sense it is a Fourier basis.

For notational clarity, in this section we will use $U_i$ to refer to the unitaries constructed from the map $\um$, and we will use $V_i$ to denote the unitaries in the $\epsilon$-local projector ${\cal P}$; i.e., $V_i$ refers to the unitaries in the $\epsilon$-local projector to which the map $\vm$ is applied.

The computation of the operator $P$ in this new basis (\ref{newbasis}) can be simplified using some inequalities.
In Eq.~(\ref{Pdef}), we defined
\be
P=\sum_{n_1,...,n_d} |n_1,...,n_d\rangle \langle n_1,...,n_d| \otimes E(\vec \theta).
\ee
However, we have also the bound that

$\Vert H_{loc}(\vec \theta) - E(\vec \theta) \Vert \leq 2^d O(\epsilon)$.  
Define
\be
\hat H_{loc} =  \sum_{n_1,...,n_d} |n_1,...,n_d\rangle \langle n_1,...,n_d| \otimes H_{loc}(\vec \theta).
\ee
Thus,
\be
\Vert P - \hat H_{loc} \Vert \leq 2^d O(\epsilon).
\ee
So, if we can compute $\hat H_{loc}$ in this new basis (\ref{newbasis}), it gives us an approximation to $\hat P$ up to small error in operator norm.

We now compute the matrix element $\langle \vec m'; a' | \hat H_{loc} | \vec m; a \rangle$.  We now define a tuple $\delta m_i$.  For given $a,a'$, define $\omega_i$ as in definition (\ref{twistdefine}).
If $\omega_i=\exp(i \theta_i)$, then set $\delta m_i=+1$.  If $\omega_i=\exp(-i \theta_i)$, then set $\delta m_i=-1$.  Otherwise, set $n_i=0$.
An explicit calculation shows that
\begin{eqnarray}
_F\langle \vec m'; a' | \hat H_{loc} | \vec m; a \rangle_F
&=&
\frac{1}{N^{D}}
\sum_{\vec n} \exp(i 2\pi \frac{\vec n \cdot(\vec m- \vec m')}{N}) \langle \vec n; a' | \hat H_{loc} | \vec n; a \rangle
\\ \nonumber
&=&
\frac{1}{N^{D}}
\sum_{\vec n} \exp(i 2\pi \frac{\vec n \cdot (\vec m-\vec m')}{N}) \exp(i 2\pi \frac{\vec n \cdot \vec{\delta m}}{N}) \exp( (v_{a'} , H_{loc} v_a)
\\ \nonumber
&=&
 \delta_{\vec m',\vec m+\vec{\delta m}}
\langle v_{a'} | H_{loc} v_a \rangle. 
\end{eqnarray}

Note that
\be
\Vert [\hat H_{loc}, U_i] \Vert = O(1/N).
\ee

So, for each $\vec m$, we have a copy of the Hilbert space of the original Hamiltonian $H_{loc}$, so in total we have $N^d$ copies of the original Hilbert space.  Let us slightly deform the $U_i$ by an amount that is $O(1/N)$.  On a copy of the Hilbert space corresponding to a given $\vec n$, and hence a given $\vec \theta=\frac{2 \pi}{N} \vec n$, we have $U_i$ equal to the identity operator times $\exp(i\frac{2\pi}{N} n_i)$.  We modify this, replacing $U_i$ with 
\be
\label{deformed}
U'_i=
\exp(i\frac{2\pi}{N} n_i) V_i^{1/N}
\ee
We choose the branch cut in $V_i^{1/N}$ in such a way that an eigenvalue of the form $\exp(i \theta)$ with $\theta\in [0,2\pi)$ is mapped to $\exp(i \theta/N)$.

This has the following interpretation: the Hamiltonian $\hat H_{loc}$ can be regarded as a ``cover" of the original Hamiltonian: we take the original $d$-dimensional torus, and tile it with $N^d$ different squares of linear size $2\pi/N$.  Then, on each square we place a copy of the Hilbert space of the original Hamiltonian $H_{loc}$, and matrix elements of the Hamiltonian can act within the same square or can connect neighboring squares.

Remark: physicists should think of each ``tile" as a ``unit cell" of some periodic lattice, and the mapping means that $\hat H_{loc}$ contains $N^d$ unit cells.

We show that $\um \circ \vm$ is close to the identity by first constructing a continuous path of strictly local Hamiltonians from $\hat H_{loc}$ to an operator which is equal to $H_{loc}$ direct summmed with another operator that exactly commutes with the $U_i$.  In this path, we change both $\hat H_{loc}$ and we also change the unitaries $U_i$ along the path.
We then use this path to construct a path of
$\epsilon'$-local projectors from $\um \circ \vm {\cal P}$ to an $\epsilon'$-local projector which is equal to ${\cal P}$ plus a trivial local projector, again changing the $U_i$ along the path.  Finally, we show that the existence of such a path gives an upper bound to the distance from $\um \circ \vm {\cal P}$ to ${\cal P}$.

\begin{lemma}
\label{pathexists}
There is a continuous path of $\pi/4$-strictly local Hamiltonians $\hat H_{loc}(s)$, with $0 \leq s \leq 1$ and of unitaries $U_i(s)$,
such that $\Vert \hat H_{loc}(s)^2-\hat H_{loc}(s) \Vert \leq 2^d O(\epsilon)$ and $\Vert [\hat H_{loc}(s),U_i(s)] \Vert \leq 2^d O(\epsilon)$ and
such that $\hat H_{loc}(0)=\hat H_{loc}$ and $\hat H_{loc}(1)$ is equal to $H_{loc}$ direct summmed with another operator that exactly commutes with the $U_i(s)$.

Let $P(s)$ be the projector onto the eigenspace of $\hat H_{loc}(s)$ with eigenvalue less than $1/2$.  For sufficiently small $\epsilon$, this is a continuous path of projectors
with $P(1)$ equal to the projector onto the eigenspace of $\hat H_{loc}(1)$ with eigenvalue less than $1/2$ direct summed with some other projector that exactly commutes with the $U_i(1)$.
Further, for all $0 \leq s \leq 1$, we have
\be
\label{PsComm}
\Vert [ P(s),U_i(s) ] \Vert \leq 2^d O(\epsilon).
\ee
\begin{proof}
We first construct $\hat H_{loc}(s)$ and only at the end of the proof do we consider $P(s)$.

After deforming the $U_i$, so that Eq.~(\ref{deformed}) holds, we have
\be
\Vert [U'_i,\hat H_{loc}] \Vert \leq \frac{\epsilon}{N}.
\ee

The next step further deforms the eigenvalues of the $U'_i$.  We do this as follows.  First, we take the first coordinate and take all of the ``tiles" with $m_1=0$, and ``stretch" them out to cover the entire torus, while deforming the other tiles so so that they all lie on the line $U_1=1$.  We then repeat this for the second coordinate, stretching the tiles with $m_2=0$, and so on, in turn for all coordinates.  After this process, one tile, with $m_1=m_2=...=0$ fills the entire torus and all other tiles are at $U_i=1$.  By ``stretching" the tile we modify it by constructing a $U''_i$ such that $U''_i=V_i$ in the given tile and $U_i=1$ elsewhere.  This deformation can be done by a smooth path of the $U_i$, and after the deformation, we have
\be
\Vert [U''_i,\hat H_{loc} \Vert \leq \epsilon.
\ee

We now construct a path from the given $U''_i$ and the given $\hat H_{loc}$ to the {\it same} $U''_i$ and to an operator that equals $H_{loc}$ acting on the tile with $m_1=m_2=...=0$ direct summed with some operator acting on the other tiles with no terms coupling the tile with $m_1=m_2=...=0$ to the other tiles.  On this path, the $U''_i$ will not change, and only the operator $\hat H_{loc}$ will change.
Once we construct this path, we are done as the resulting operator is equal to the original $H_{loc}$ up to direct summing with an operator that commutes with all the $U_i$.

To construct this path, we consider a slightly different notation to describe the same situation.  Define operators $M_i$ such that $M_i=m_i$ in a given tile.  Let $W_i$ be unitaries which are block-diagonal, where the blocks correspnd to the tiles, with $W_i$ equal to $U_i$ in every tile.
Then, unitaries $W_i$ and operator $\hat H_{loc}$ are such that every eigenvectors of the $W_i$ is at least $N^d$-fold degenerate (if the $U_i$ have a degeneracy in their eigenvalues, then the $W_i$ have a degeneracy equal to $N^d$ times the degeneracy of the $U_i$), with the degeneracy corresponding to different choices of the $M_i$.  The operator $\hat H_{loc}$ is $R$-strictly local with respect to the $W_i$, however it does not commute with $M_i$.  Instead, the operator $\hat H_{loc}$ has matrix elements that can increase or decrease $M_i$ by one.
Let $\tilde M_i$ denote the operator that increases $M_i$ by $1$, mod $N$.  Then, $\hat H_{loc}$ commutes with $\tilde M_i$.  The operator $\tilde M_i$ has eigenvalues which are $N$-th roots of unity.  What we will do is to define an operator $\hat H_{loc}(s)$, for $s$ varying from $0$ to $1$, with $\hat H_{loc}(0)=\hat H_{loc}$.
 The operator $\hat H_{loc}$ commutes with the $\tilde M_i$, so it can be written as a block-diagonal matrix, where now the blocks correspond to the different eigenspaces of the $\tilde M_i$ with different.  The eigenvalues of the $\tilde M_i$ are $N$-th roots of unity, so there are $N^d$ blocks.
We define $\hat H_{loc}(s)$ by twisting
the operators in each of the $N^d$ different blocks by some angle,
using the operators $W_i$ as the unitaries used to define the twist.  In the eigenspace of the $\tilde M_i$ with eigenvalues $\tilde m_i=\exp(i \phi_i)$ with $\phi_0 \in [0,2\pi)$, we twist by angle
\be
s \phi_i.
\ee
To understand the effect of this twist, consider a matrix element of $ H_{loc}$ which increases $M_i$ by one.  Call this matrix element $(H_{loc})_{ab}$, where $a,b$ label particular basis vectors.  In the eigenspace of $\tilde M_i$ with eigenvalue $\tilde m_i$, the corresponding matrix element of $\hat H_{loc}$ is $\tilde m_i (H_{loc})_{ab}$.  Conversely, if the matrix element of $H_{loc}$ decreases $M_i$ by one, then the corresponding matrix 
element of $\hat H_{loc}$ is $(\tilde m_i^{-1} H_{loc})_{ab}$.
The effect of the twist is to cancel this factor of $\tilde m_i$ or $\tilde m_i^{-1}$ so that $H_{loc}(1)$ has the same matrix elements  as $H_{loc}$.
So, $\hat H_{loc}(1)$ is equal to the direct sum of $N^d$ different copies of $H_{loc}$.  The operator $\hat H_{loc}(1)$ commutes with all the $M_i$, and is equivalent to $H_{loc}$ in each copy.

Now, we use this same path $\hat H_{loc}(s)$ to define the path from our original $\hat H_{loc}$ to the desired final operator which has no terms coupling the tile $m_1=m_2=...=0$ to other tiles.

So, it remains to show that $H_{loc}(s)$ is close to a projector for all $0 \leq s \leq 1$ and that it almost commutes with the $U''_i$ for all such $s$.
However, both these follow from lemma (\ref{HIsAppProj}) and lemma (\ref{twistedAlmostCommutes}), so it is within $2^d O(\epsilon$ of a projector and its commutator with the $U''_i$ is bounded by $2^d O(\epsilon )$.

Using the bound on $\Vert H_{loc}(s)^2-\hat H_{loc}(s) \Vert$, we can bound $\Vert \hat H_{loc}(s) - P(s) \Vert \leq 2^d O(\epsilon)$.  Using this bound and the bound on
$\Vert [\hat H_{loc}(s),U_i] \Vert$, Eq.~(\ref{PsComm}) follows.
\end{proof}
\end{lemma}

\begin{lemma}
\label{Ppath}
Let ${\cal P}(s)$ be a family of $\epsilon$-local projectors which depends continuously upon a parameter $s$ with $0\leq s \leq 1$. Denote the projector corresponding to ${\cal P}(s)$ by $P(s)$ and denote the unitaries along the path by $U_i(s)$.
Then,
\be
{\rm dist}({\cal P}(0),{\cal P}(1)) \leq O(\epsilon).
\ee
\begin{proof}
Let $M$ be a large integer chosen later.
We define new $\epsilon$-local projectors ${\cal P}'(0)$ and ${\cal P}'(1)$.
In ${\cal P}'(0)$,
we replace the unitary matrices $U_i$ with $U'_i(0) \equiv U_i(0) \oplus U_i(\frac{1}{M}) \oplus U_i(\frac{1}{M}) \oplus U_i(\frac{2}{M}) \oplus U_i(\frac{2}{M}) \oplus ... \oplus U_i(\frac{M-1}{M}) \oplus U_i(\frac{M-1}{M})$, so that there are a total of $2M-1$ copies of the original $U_i$.
In ${\cal P}'(1)$,
we replace the unitary matrices $U_i$ with $U'_i(1) \equiv U_i(\frac{1}{M}) \oplus U_i(\frac{1}{M}) \oplus U_i(\frac{2}{M}) \oplus U_i(\frac{2}{M}) \oplus ... \oplus U_i(\frac{M-1}{M}) \oplus U_i(\frac{M-1}{M}) \oplus U_i(1)$, so that there are again a total of $2M-1$ copies of the original $U_i$.  Note that $\Vert U'_i(0) - U'_i(1) \Vert$ can be made arbitrarily small by taking $M$ large.

In ${\cal P}'(0)$, we replace $P(0)$ by
$P'(0)\equiv P(0)\oplus I \oplus 0 \oplus I \oplus 0 ... \oplus I \oplus 0$, adding a total of $M-1$ copies of the identity matrix and $M-1$ copies of the zero matrix.  In ${\cal P}'(1)$, we replace $P(1)$ by
$P'(1)\equiv P(1) \oplus I \oplus 0 \oplus I \oplus 0 ... \oplus I \oplus 0$ similarly.  

We regard this matrix $P(0)$ as a block-diagonal matrix consisting of $M$ blocks, with the first block equaling $P(0)$, and the other blocks equaling $I \oplus 0$.
Label the blocks by $b=0,...,M-1$.
We make a basis change; this basis change does not change the block $b=0$, but replacing $I \oplus 0$ in block $b>0$ with $P(\frac{b}{M}) \oplus (I-P(\frac{b}{M}))$.
We make a similar basis change in $P(1)$, replacing $I \oplus 0$ in block $b>0$ with $P((b)/M) \oplus (I-P((b)/M))$ in the same way.

Having made this basis change, we now compute $\gm {\cal P}'(0)$.  Making a basis change, we can write $P'(0) U'_i P'(0)$ as 
$P(0) U_i(0) P(0)
 \oplus P(\frac{1}{M}) U_i(1/m) P(\frac{1}{M}) \oplus (1-P(\frac{1}{M})) U_i(\frac{1}{M}) (1-P(\frac{1}{M})) \oplus ...
\oplus  P(\frac{M-1}{M}) U_i(\frac{M-1}{M}) P(\frac{M-1}{M}) \oplus (1-P(\frac{M-1}{M})) U_i(\frac{M-1}{M}) (1-P(\frac{M-1}{M}))$.  We write this more
compactly as
$$P(0) U_i(0) P(0) \oplus \oplus_{b=1}^{M-1}  \Bigl( P(\frac{b}{M}) U_i(\frac{b}{M}) P(\frac{b}{M}) \oplus (1-P(\frac{b}{M})) U_i(\frac{b}{M}) (1-P(\frac{b}{M})\Bigr).$$
Similarly, after a different basis change, we can write
$P'(1) U'_i P'(1)$ as 
$$ \oplus_{b=1}^{M-1}  \Bigl( P(\frac{b}{M}) U_i(\frac{b}{M}) P(\frac{b}{M}) \oplus (1-P(\frac{b}{M})) U_i(\frac{b}{M}) (1-P(\frac{b}{M})\Bigr) \oplus
P(1) U_i(1) P(1).$$

Making a further basis change to re-order these we can write $P'(0) U'_i P'(0)$ as $$
\Bigl( \oplus_{b=0}^{M-1} P(\frac{b}{M}) U_i(\frac{b}{M}) P(\frac{b}{M}) \Bigr) \oplus \Bigl( \oplus_{b=1}^{M-1}  (1-P(\frac{b}{M})) U_i(\frac{b}{M}) (1-P(\frac{b}{M})\Bigr)$$ and write
$P'(1) U'_i P'(1)$ as 
$$
\Bigl( \oplus_{b=1}^{M} P(\frac{b}{M}) U_i(\frac{b}{M}) P(\frac{b}{M}) \Bigr) \oplus \Bigl( \oplus_{b=1}^{M-1}  (1-P(\frac{b}{M})) U_i(\frac{b}{M}) (1-P(\frac{b}{M})\Bigr)$$

The distance between these is bounded by ${\rm max}_{b\in \{0,...,M-1\}} \Vert P(\frac{b}{M}) U_i(\frac{b}{M}) P(\frac{b}{M}) - P(\frac{b+1}{M}) U_i(\frac{b+1}{M}) P(\frac{b+1}{M}) \Vert$, 
which can be made arbitrarily small by taking $M$ large, using the continuity assumption on the path.
The unitaries in $\gm {\cal P}'(0)$ and $\gm {\cal P}'(1)$ are given by taking the polar of these $P'(0) U'_i(0) P'(0)$ and $P'(1) U'_i(1) P'(1)$ and so we we can makethe distance between these unitaries arbitrarily small.  So, by lemma \ref{swindle}, the desired result follows.
\end{proof}
\end{lemma}

\begin{lemma}
\label{umvm}
Let ${\cal P}$ be an $\epsilon$-local projector.  Then,
${\rm dist}({\cal P},\um \circ \vm {\cal P})= 2^d O(\epsilon)$, where we pick $N=1/\epsilon$ to define the map $\um$.
\begin{proof}
This follows from lemmas \ref{pathexists},\ref{Ppath}.
\end{proof}
\end{lemma}

\subsection{The Composition of Maps $\vm\circ\um$}
We now compute the composition of maps $\vm\circ\um$, mapping a projector-valued function $E$, to an $\epsilon$-local projector, and back to a projector-valued function.

Let the projector $E(\vec \theta)$ have rank $r$ and act on a space of dimension $D$.  We consider such a function $E$ to be trivial if there exists a family $E(\vec \theta,s)$ of projectors, with $s\in [0,1]$, such that $E(\vec \theta,0)=E(\vec \theta)$ and such that $E(\vec \theta,1)$ is a constant function, independent of $\theta$, and such that
$E(\vec \theta,s)$ is an infinitely differentiable function of $\theta$ and $s$.  This definition is equivalent to defining a function $E(\vec \theta)$ to be trivial
if we can define an isometry, $A(\vec \theta)$, such that $E(\vec \theta)=A(\vec \theta) A(\vec \theta)^\dagger$ and such that $A$ is infinitely differentiable.
Finally, this implies that a function $E(\vec \theta)$ is trivial if we there exists a family $E(\vec \theta,s)$ of projectors, so $s\in [0,1]$, such that $E(\vec \theta,0)=E(\vec \theta)$ and such that $E(\vec \theta,1)$ is some other trivial function, independent of $\theta$, and such that
$E(\vec \theta,s)$ is an infinitely differentiable function of $\theta$ and $s$.

Finally, note that if $F$ is trivial and if $\Vert F(\vec \theta)-E(\vec \theta) \Vert$ is sufficiently small, then $E$ is trivial.  To prove this, consider the operator
$O(\vec \theta,s)=s F(\vec \theta) + (1-s) E(\vec \theta)$.  For a sufficiently small bound on $\Vert F(\vec \theta)-E(\vec \theta) \Vert$, we can uniformly bound the eigenvalues of $O(\vec \theta,s)$ away from $1/2$ for $0 \leq s \leq 1$, and so the projector onto the eigenspace of $O(\vec \theta,s)$ with eigenvalue greater than $1/2$ is infinitely differentiable and defines a family $E(\vec \theta,s)$ of projectors as above.

\begin{lemma}
\label{trivn}
Consider a projector-valued function $E$ obeying Eq.~(\ref{Lipschitz}).  For any given $K$ in Eq.~(\ref{Lipschitz}), for all sufficiently large $N$ and for all sufficiently small $\delta>0$, if applying the map $\um$ to this projector gives an $\epsilon$-local projector ${\cal P}$ such that ${\cal P}$ has distance
at most $\delta$ from some trivial local projector, then
$E$ has the property that $\oplus_{i=1}^{N^d} E$ is trivial.
Here, $\oplus_{i=1}^{N^d} E$ denotes the function that maps angles $\vec \theta$ to the direct sum of $N^d$ copies of $E(\vec \theta)$.
\begin{proof}
Let $F=\oplus_{i=1}^{N^d} E$.
Let $E(\vec \theta)$ have rank $r$, so that $F(\vec \theta)$ has rank $r N^d$.

Define $\hat P$ and $U_i$ as before from $E$.
By assumption, we can add some trivial projector ${\cal R}$ to $\hat P$ to obtain an $\epsilon$-local projector that has distance at most $\delta$ from some trivial local projector.  Let us deform all the unitaries in both trivial local projectors so that their eigenvalues are $N$-th roots of unity.
This can be done whlie keeping the property that the unitaries commutes with each other and commute with the projector (i.e., while maintaining the property of being a trivial local projector), and changes the unitaries only by $O(1/N)$.  So, this change in the trivial local projectors $V_i$ increases the distance from ${\cal P}+{\cal R}$ to an trivial projector by at most $O(1/N)$, so that the distance is $\delta+O(1/N)$.
Having done this, the projector ${\cal R}$ is in contained in the image under $\um$ of some projector-valued function $G$ which has the property that $G(\theta)$ is independent of $\theta$.  Note that $\um G$ is trivial, so $\um (E+G)$ is within distance at most $\delta+O(1/N)$ from some trivial projector ${\cal Q}$.  Let ${\cal Q}$ have projector $Q$ and unitaries $V_i$.

Let $L=E+G$ and let $L(\vec \theta)$ have rank $l$.

For integers $m_1,...,m_d$, define $L_{\vec m}$ so that $L_{\vec m}(\vec \theta)=L(\vec \theta+2 \pi \frac{\vec m}{N})$.
Define $M=\oplus_{\vec m} L_{\vec m}$, where the sum is over $0\leq m_1,...,m_d < N$.
Note that $M$ can be smoothly deformed to $\oplus_{i=1}^{N^d}  L$.  So, if we show that $M$ is trivial, it will imply that
$\oplus_{i=1}^{N^d} L$ and $\oplus_{i=1}^{N^d} E$ are trivial, as desired.

Let $\hat M$
be defined by
$\hat M=\sum_{n_1,...,n_d} |n_1,...,n_d\rangle \langle n_1,...,n_d| \otimes M(2\pi \frac{n_1}{N},...,2\pi \frac{n_d}{N})$.
Note that $M(\vec \theta)$ is a block diagonal matrix with $N^d$ blocks, each block being labeled by integers $m_1,...,m_d$.
Define the operator $\Delta_{\vec {\delta m}}$ as follows; this operator is a matrix of the same size as $M$ and it permutes the blocks, sending the block labeled with integers $\vec m$ to the block with integers labeled by $\vec m + \vec \delta m$.
Define $\hat \Delta_{\vec {\delta m}}=\sum_{n_1,...,n_d} |n_1,...,n_d\rangle \langle n_1,...,n_d| \otimes \Delta_{\vec {\delta m}}$.

Define $\hat Q$ by $\hat Q=\sum_{n_1,...,n_d} |n_1,...,n_d\rangle \langle n_1,...,n_d| \otimes (\Delta_{-\vec n} Q \Delta_{\vec n})$.  Note that $\hat M$ and $\hat Q$ are matrices of the same size.
Define unitaries $\hat V_i$ as follows.  These will be unitaries of the same dimension as $\hat M$ and $\hat Q$.
Then, let
\be
\hat V_i=\sum_{n_1,...,n_d} |\vec n + \vec x(i) \rangle \langle \vec n |
\otimes \Delta_{-\vec x(i)} V_i.
\ee
Note that
\be
\Vert \Delta_{-\vec x(i)} V_i -I \Vert \leq \delta+O(1/N),
\ee
because the unitaries $V_i$ are within distance $\delta+O(1/N)$ of the unitaries in $\um (E+G)$, and those unitaries in $\um (E+G)$ are precisely equal to $\Delta_{-\vec x(i)}$.
So, the resulting $\hat V_i$ are close to the unitaries in $\um M$ and $\hat Q$ is close to $\hat M$ and
\be
[\hat V_i,\hat Q]=0
\ee
by construction.

So, pick any basis for the range of $Q$.  By applying  $\Delta_{-\vec x(i)} V_i$, we  obtain a basis for the range of $\Delta_{-\vec x(i)} Q \Delta_{\vec x(i)}$.  By applying a sequence of  $\Delta_{-\vec x(i)} V_i$ for various choices of $i$, we can obtain a basis for the range of $\Delta_{-\vec n} Q \Delta_{\vec n}$ for any $n$, where $n$ is equal to the sum of $\vec x(i)$ along the sequence; we can regard this sequence as a ``path" from the zero vector to $\vec n$.  Because the $\hat V_i$ commute with each other, this basis is independent of the choice of path.
 Corresponding to this choice of basis is an isometry that we write $A(\vec \theta)$ for $\vec \theta = 2 \pi \frac{\vec n}{N}$.  This isometry is from an $l N^d$-dimensional space to the range of $\Delta_{-\vec n} Q \Delta_{\vec n}$, with the property that 
\be
\label{LA1}
\frac{\Vert A(\vec \theta)-A(\vec \theta') \Vert}{{\rm dist}(\vec \theta,\vec \theta')} \leq N \epsilon (\delta+O(1/N)),
\ee
where we use the fact that the $\hat V_i$ are close to the identity to obtain this Lipshitz condition.
We now extend this to an isometry $A(\vec \theta)$ for {\it all} $\theta$, with $A(\vec \theta)$ obeying a Lipschitz condition.
Eq.~(\ref{LA1}) will play a key role in this extension.

To extend the isometry, we write the torus as the union of $N^d$ hypercubes.  Each hypercube is labeled by integers $n_1,...,n_d$ and contains the points
such that $2\pi \frac{n_i}{N} \leq \theta \leq 2\pi\frac{n_i+1}{N}$ for all $i$, treating the quantities periodic mod $2\pi$ in the natural way.  The boundaries of hypercubes overlap.  We extend $A(\vec \theta)$ to some isometry on each hypercube so that the extension is consistent on the boundaries.
The strategy to define this is to first extend $A(\vec \theta)$ to an approximate isometry by a simple interpolation procedure, and then to approximate the 
approximate isometry by an exact isometry.  We now give one possible implementation of this strategy to
define the extension on a given hypercube; other implementations are possible.  We have fixed $A(\vec \theta)$ on the vertices of the hypercube.  Label the vertices of a hypercube by integers $b_1,...,b_d$ with $b_i \in \{0,1\}$, so that a vertex with a given set of $b_i$ is at coordinate
$\theta_i=2\pi \frac{n_i+b_i}{N}$.  For any point $\vec \theta$ in any given hypercube labeled by integers $\vec n$, define real numbers $x_i \in [0,1]$ by $x_i=\frac{N}{2\pi}\theta_i - n_i$.
Then let
\be
B(\vec \theta)=\sum_{\{b_i\}} \prod_i \Bigl(x_i b_i + (1-x_i) (1-b_i)\Bigr) A(2\pi \frac{\vec n + \vec b}{N}).
\ee
Then, for $\vec \theta$ equal to a vertex of the hypercube, we have $B(\vec \theta)=A(\vec \theta)$.  If $\vec \theta$ is in more than one hypercube, this definition of $B(\vec \theta)$ is independent of which hypercube we pick to define $B(\vec \theta)$.
By construction, $B(\vec \theta)$ obeys a Lipschitz condition:
\be
\label{LipschitzB}
\frac{\Vert B(\vec \theta)-B(\vec \theta') \Vert}{{\rm dist}(\vec \theta,\vec \theta')} \leq N \epsilon (\delta+O(1/N)).
\ee
The resulting $B$ need not be a differentiable function of $\vec \theta$, but we can construct an infinitely differentiable function by convolving $B$ by a infinitely differentiable function whose support extends a distance $O(1/N)$ from the origin.  Choosing this function we convolve with to be positive and integrate to unity, the resulting convolution gives an infinitely differentiable function that obeys a similar condition of Eq.~(\ref{LipschitzB}) and the resulting infinitely differentiable function of $\vec \theta$ is close to $B$.  So, if one desires to construct a smooth projector-valued function in this lemma, then from here on, when we refer to $B$, one can assume we mean $B$ convolved in this way.

The matrix $B(\vec \theta)$ is not necessarily an isometry, but it is an approximate isometry because $\Vert A(\vec \theta_1)-A(\vec \theta_2) \Vert$ is $O(\epsilon)$.  That is, $B(\vec \theta)^\dagger B(\vec \theta)$ is approximately equal to the identity matrix.
Define for arbitrary $\vec \theta$ that
\be
A(\vec \theta)=B(\vec \theta) \Bigl( B(\vec \theta)^\dagger B(\vec \theta) \Bigr)^{-1/2}.
\ee
This definition is similar to the definition of a polar of a matrix.  If we pick $\epsilon$ sufficiently small that $\Vert B(\vec \theta)^\dagger B(\vec \theta)-I \Vert$ is bounded by some constant strictly less than one,
 then $A(\vec \theta)$ also obeys a Lipschitz condition:
\be
\frac{\Vert A(\vec \theta)-A(\vec \theta') \Vert}{{\rm dist}(\vec \theta,\vec \theta')} \leq N O(\epsilon) (\delta+O(1/N)).
\ee

Having defined $A(\vec \theta)$, note that $\Delta_{-\vec n} Q \Delta_{\vec n}=A(\vec \theta) A(\vec \theta)^\dagger$ for $\vec \theta$ on the vertex of a hypercube, and so by the bound on $\Vert M(\vec \theta)-\Delta_{-\vec n} Q \Delta_{\vec n} \Vert$ and by the Lipschitz conditions, we have
\be
\label{gotit}
\Vert M(\vec \theta)-A(\vec \theta) A(\vec \theta)^\dagger \Vert \leq \delta + O(1/N).
\ee
Let $M'(\vec \theta)=A(\vec \theta) A(\vec \theta)^\dagger$.  By construction, $M'$ is trivial, and so by Eq.~(\ref{gotit}),
$M$ is trivial for sufficiently small $\delta$ and sufficiently large $N$.
\end{proof}
\end{lemma}

We will use the above lemma to
show that for sufficiently large $N$, $\oplus_{i=1}^{N^d} \Bigl(\um \circ \vm E-E\Bigr )$ is trivial in lemma (\ref{absnon}) below.  Before doing this, we need some more definitions of sum, difference, and inverse.
In definition (\ref{sum}), we have defined the sum of two $\epsilon$-local projectors.  We now define an inverse and a difference.
\begin{definition}
Given an $\epsilon$-local projector ${\cal P}$, corresponding to projector $P$ and unitaries $U_1,...,U_d$, let $\overline{\cal P}$ denote the $\epsilon$-local projector corresponding to projector $1-P$ and unitaries $U_1,...,U_d$.

Given two $\epsilon$-local projectors, ${\cal P},{\cal Q}$, define ${\cal P}-{\cal Q}={\cal P}+\overline {\cal Q}$.
\end{definition}

\begin{lemma}
For any $\epsilon$-local projector ${\cal P}$, the $\epsilon$-local projector ${\cal P}+\overline {\cal P}$ is within distance $O(\epsilon)$ of a trivial local projector.
\begin{proof}
Let ${\cal P}$ correspond to projector $P$ and unitaries $U_1,...,U_d$.
We will write all the matrices in ${\cal P}+\overline {\cal P}$ as $4$-by-$4$ block matrices, with the first two blocks corresponding to the matrices in ${\cal P}$ and the second two blocks corresponding to the matrices in $\overline {\cal P}$.  The first and third blocks will be the range of $P$.  Then, the projector in ${\cal P}+\overline {\cal P}$ equals
\be
\label{projdiff}
\begin{pmatrix} I &&& \\ &0&& \\ &&0&\\&&&I\end{pmatrix},
\ee
and the unitaries in ${\cal P}+\overline {\cal P}$ equal
\be
\begin{pmatrix} U_i^{11} & U_i^{12} && \\ U_i^{21} & U_i^{22} && \\ && U_i^{11} & U_i^{12}  \\ && U_i^{21} & U_i^{22} \end{pmatrix}.
\ee
We approximate the matrix in the above equation by
\be
\label{uapprox}
\begin{pmatrix} U_i^{11} &&& U_i^{12}  \\  & U_i^{22} & U_i^{21} & \\ & U_i^{12} & U_i^{11}&  \\ U_i^{21} &&& U_i^{22} \end{pmatrix}.
\ee
\end{proof}
The unitaries in Eq.~(\ref{uapprox}) exactly commute with each other and with the projector in Eq.~(\ref{projdiff}), so they form a trivial local projector.  Given the bound that
$\Vert U_i^{12} \Vert \leq \Vert [P,U_i] \Vert$, the result follows.
\end{lemma}

\begin{definition}
Given any projector valued function $E$, we define $\overline E$ so that $\overline E(\vec \theta)=I-E(\vec \theta)$.
We similarly define a sum and difference of projectors $E,F$ so that $E+F$ denotes the projector-valued function with $(E+F)(\vec \theta)=E(\vec \theta)+F(\vec \theta)$ and $E-F$ denotes $E+\overline F$.
\end{definition}

\begin{lemma}
\label{absnon}
Consider any projector-valued function $E$.  Then for sufficiently large $N$, $\oplus_{i=1}^{N^d} \Bigl(\vm \circ \um E-E\Bigr )$ is trivial.
\begin{proof}
Choose $N>1/\epsilon$.
By lemma \ref{umvm}, applied to ${\cal P}=\um E$, we have ${\rm dist}(\um E,\um \circ \vm \circ \um E)=2^d O(\epsilon)$.  Hence, $\um E-\um \circ \vm \circ \um E$ is within distance $2^d O(\epsilon)$ of a trivial local projector.

Note that $-\um \circ \vm \circ \um E$ is the image under $\um$ of $\vm \circ \um \overline E$.  So, the image under $\um$ of $E+\vm \circ \um \overline E$ is trivial and so by lemma (\ref{trivn}), $\oplus_{i=1}^{N^d} \Bigl(\vm \circ \um E-E\Bigr )$ is trivial.
\end{proof}
\end{lemma}

Finally, we have
\begin{lemma}
\label{classify}
Consider any projector-valued function $E$.  Then for sufficiently large odd $N$, $\vm \circ \um E-E$ is trivial.
\begin{proof}
The $E$ are classified by invariants which are either integers or are in $Z_2$.  The $Z_2$ invariants occur only for the cases with symmetries as considered in section \ref{symmsec}, while in the cases considered thus far we have only integer invariants.  In all of these cases, if $N$ is odd then if $\oplus_{i=1}^{N^d} F$ is trivial for any $F$ then $F$ is trivial.  Take $F=\vm \circ \um E-E$ and apply lemma \ref{absnon}.


Note that if the invariants are only integer invariants and not $Z_2$, then the restriction to odd $N$ is not necessary.
\end{proof}
\end{lemma}

\section{Example}
In this final section, we present an interesting example of a {\it similar} procedure to that described above.  In fact, in this section we will construct a pair of almost commuting matrices directly from a line bundle without going through the intermediate steps of constructing projectors.  This pair of matrices will be equivalent, up to conjugation by a unitary, to the example of Ref.~\onlinecite{voiculescu}.
The construction in this section should be regarded as motivational for the more general approach we used previously.

For an integer $N$, we define the following natural discrete analogue of a constant curvature connection for a line bundle on the two torus.
For each pair of integers, $m,n$, we define two unitaries, $U_x(m,n)$ and $U_y(m,n)$.  These unitaries, $U_x(m,n)$ and $U_y(m,n)$ will depend periodically upon $m,n$ with period $N$.
The definitions below are all for $0 \leq m,n \leq N-1$, and for arbitrary $m,n$ one must use the periodicity.
Define $U_x(m,n)$ by
\begin{eqnarray}
m \neq N-1  &\; \rightarrow \;  & U_x(m,n)=1, \\ \nonumber
m=N-1  & \; \rightarrow \; & U_x(m,n)=\exp(-i 2\pi \frac{n}{N}),
\end{eqnarray}
and define $U_y(m,n)$ by
\begin{eqnarray}
 U_y(m,n)=\exp(i 2\pi \frac{m}{N^2}).
\end{eqnarray}
We heuristically regard $U_x(m,n)$ as phase resulting from transport from $m,n$ to $m+1,n$ and $U_y(m,n)$ as a phase resulting from transport from $m,n$ to $m,n+1$.
Similarly, we regard $U_x(m,n)^\dagger$ as phase resulting from transport from $m,n$ to $m-1,n$ and $U_y(m,n)^\dagger$ as a phase resulting from transport from $m,n$ to $m,n-1$.

Compute now the product of phases resulting from transport around a single square starting at $m,n$ and moving right, up, left, down in sequence.
This phase is $U_y(m,n+1)^\dagger U_x(m+1,n+1)^\dagger U_y(m+1,n) U_x(m,n)$.
A calculation gives that this phase is $\exp(i 2\pi \frac{1}{N^2})$, independent of $m,n$, reminiscent of the constant curvature.
Note that $U_x(N-1,N-1)$ is not close to $U_x(N-1,N)$; this is analogous to a singularity in the connection.

Define an $N^2$ dimensional space with basis elements written $|m,n\rangle$, with $m,n$ periodic in $N$.
Now consider the pair of unitaries given by
\begin{eqnarray}
U&=&\sum_{m,n} U_x(m,n) |m+1,n\rangle \langle m,n|, \\ \nonumber
V&=&\sum_{m,n} U_y(m,n) |m,n+1\rangle \langle m,n|.
\end{eqnarray}
These definitions are reminiscent of the definitions in Eq.~(\ref{UiDef1}).  The difference here is that the phase $U_x(m,n)$ or $U_y(m,n)$ appears in the definition.

A computation shows that $V^\dagger U^\dagger V U = \exp(i 2\pi \frac{1}{N^2})$.
Further, a calculation shows that both $U$ and $V$ have the set of eigenvalues $\exp(i 2\pi \frac{j}{N^2})$ for $j=0,...,N^2-1$.
Thus, up to a basis change, these unitaries are the same as the original Voiculescu unitaries.

It might be interesting to consider generalizing this procedure in this section to more general bundles and connections.

\section{Symmetries}
\label{symmsec}
In applications in physics, often one considers systems with various symmetries such as time-reversal symmetry, particle-hole symmetry, or chiral symmetry.  These symmetries play an important role in a classification of possible phases of topological insulators\cite{kitaev,topo10}.  A total of 10 such symmetry classes have been discovered, analogous to the 10-fold way in random matrix theory\cite{10rmt}.  These 10 classes break into two different complex classes and eight different real classes, corresponding to two-fold or eight-fold Bott periodicity\cite{kitaev}.

In Ref.~\onlinecite{HL1,HL2}, classification of almos-commuting self-dual unitary matrices was applied to disordered time-reversal invariant topological insulators.  This is a start of extending the classification of almost-commuting matrices with these symmetries.

In this section, we present some brief comments on how the results above can be extended to the case of symmetries.  We consider only two cases (we have already considered above the case in which the matrices $U_i$ are arbitrary unitaries, the projector $P$ is an arbitrary projector, and the projector-valued function $E$ is arbitrary).  These two cases correspond physically to having time-reversal symmetry but not spin-orbit coupling or having time-reversal symmetry and strong spin-orbit coupling.
After explaining these two cases, we will briefly mention the other seven cases.

We refer to these two cases as the ``symmetric" and ``self-dual" cases respectively.
In the symmetric case, we impose that the matrices $U_i$ in a soft torus are symmetric.  That is,
\be
U_i=U_i^T,
\ee
where the superscript $T$ denotes the tranpose.
If the matrices in a soft torus have this property, then we refer to it as a symmetric soft torus.  If the unitaries {\it and} the projector in a local projector are symmetric, then we refer to it as a symmetric projector.

In the self-dual case, the $U_i$ in a soft torus have even size and we require that they are ``self-dual".  This means that we require that
\be
U_i=-Z U_i^T Z,
\ee
where $Z$ is an anti-symmetric block-off-diagonal matrix:
\be
Z=\begin{pmatrix} 0 & I \\ -I & 0 \end{pmatrix}.
\ee
If the matrices in a soft torus have this property, then we refer to it as a self-dual soft torus.  If the unitaries {\it and} the projector in a local projector are self-dual, then we refer to it as a self-dual projector.

We claim that the maps we have given preserve these symmetries in that the map $\fm$ maps symmetric soft tori to symmetric local projectors and self-dual soft tori to self-dual local projectors, while the map $\gm$ maps symmetric local projectors to symmetric soft tori and self-dual local projectors to self-dual soft tori.
Checking this proprty of $\gm$ is immediate.  To verify this property of $\fm$, note that the $X_i,Y_i$ inherit the symmetry of $U_i$.
We firstverify that the POVM produced inherits the symmetry (i.e., they are symmetric or self-dual, respectively).  Note that the $F(\frac{X_i}{\Delta}+m_i)$ and $F(\frac{Y_i}{\Delta}+m_i)$ inherit the symmetry of $X_i$ and $Y_i$ respectively, which in turn inherit the symmetry of the $U_i$.  One can verify then that the operators $E_{\vec m,\vec n}$ in Eq.~(\ref{Edef}) inherit the symmetry.
The key step is then to show that the operators $Q_{\vec m,\vec n}$ and $\Pi$ can be chosen to inherit the symmetry (there is some arbitrariness in the $Q_{\vec m,\vec n},\Pi$, so we simply show that a choice that inherits the symmetry exists).  In the self-dual case, this requires defining a matrix $Z$ on the larger space.  We choose this matrix $Z$ on the larger space in some arbitrary way, so that the matrix $Z$ on the larger space is equal to the matrix $Z$ on the smaller space direct summer with some other matrix.  In an abuse of notation, we use $Z$ for both matrices.
Now, we show that
\begin{lemma}
Given a POVM $E_i$ where the $E_i$ are either symmetric or self-dual, it is possible to define projectors $Q_i$ and $\Pi$ on some larger space so that
$\Pi Q_i \Pi$ is equal to $E_i$ when restricted to the range of $\Pi$ and so that $\Pi$ and $Q_i$ inherit the symmetry of $E_i$.
\begin{proof}
Note that each $E_i$ can be written as a sum $\sum_a \lambda_a P_{a,i}$, where $P_{a,i}$ projects onto the eigenspace of $E_i$ with eigenvalue $\lambda_a$.  The $P_{a,i}$ inherit the symmetry of $E_i$.

Now we claim that a symmetric projector $P$ can be written as a sum of symmetric rank-$1$ projectors and a self-dual projector can be written as a sum of self-dual rank-$2$ projectors.  In the symmetric case, suppose $P$ has non-zero rank.  Choose any real vector $v$ in the range of $P$, and write $P=|v\rangle\langle v|+(P-|v\rangle\langle v|)$.  The term in parenthesis is a projector with lower rank than $P$.  If this rank is non-zero then, repeat the procedure.  Continue until the the term in parentheses is zero, giving the desired decomposition of $P$.  Note that in the real case, the sum of rank-$1$ projectors is a sum of projectors $|v\rangle\langle v|$, where $v$ is real.  In the self-dual case, let $v$ be any vector in the rank of $P$.  Let $w=Z \overline v$, where $\overline v$ is the vector with the entries of $v$ complex conjugated.  Then, $|v \rangle \langle v| + |w \rangle \langle w|$ is self-dual and one can verify that $v,w$ are orthogonal so that this is a rank-$2$ projector.  Subtract this rank-$2$ projector from $P$.  If the result has non-zero rank, then repeat this procedure.  Continue until a zero result is obtained, giving the desired decomposition.

Write $E_i=\sum_a \lambda_a P_{a,i}$, and then further decompose the projector $P$ as a sum as in the above paragraph.
Suppose that $P_{a,i}=\sum_b |v_{b,a,i} \rangle \langle v_{b,a,i}|$ for some vectors $v$.  So, we can write
$E_i=\sum_a \sum_b |\sqrt{\lambda_a} v_{b,a,i} \rangle \langle \sqrt{\lambda_a} v_{b,a,i} |$.
Writing $\sqrt{\lambda_a} v_{b,a,i}=w_{b,a,i}$, and then combining the two indices $b,a$ into a single index we can decompose
\be
E_i=\sum_{c=1}^{N_i} | w_{c,i} \rangle \langle w_{c,i} |,
\ee
where $N_i$ is the total number of vectors in the decomposition.
In the symmetric case, the $w_{c,i}$ are real vectors.  In the self-dual case, $N_i$ is even and we can order the vectors so that
$w_{c,i}=Z \overline w_{c+1,i}$ for $c=1,3,5,...$.  

Let the index $i$ in $E_i$ range from $1$ up to $C$ from some given $C$.
 Let the original space have dimension $D$.
We define the dimension of the larger space to be $D'=\sum_{i=1}^C N_i$. Consider the sequence of vectors $w_{1,1},w_{2,1},...,w_{N_1,1},w_{1,2},w_{2,2},...,w_{N_2,2},...,w_{1,C},w_{2,C},...,w_{N_C,C}$.  Let $w_a$ be the $a$-th vector in this sequence.
Note that in the self-dual case, if $a$ is odd then $w_a=Z \overline w_{a+1}$.
Now, we define a matrix $A$ with dimension $D$-by-$D'$.  In the $a$-th column, the entires of this matrix $M$ are given by the entries of the vector $w_a$, with the $j$-th coordinate in the $j$-th row.

Note that $A A^\dagger=I$, by the fact that $\sum_i E_i=I$.  Thus, $A$ is an isometry.  Therefore, the rows of $A$ are orthonormal to each other.
Let $M$ be a matrix with dimension $D'$-by-$D'$.
We will pick the first $D$ rows of $M$ to equal the matrix $A$.  We now describe how to choose the remaining $D'-D$ rows of $M$.
The goal is to fill in the remaining rows so that $M$ is a unitary matrix and so that certain symmetries hold.  Once we have a unitary matrix $M$, then we let
$Q_i$ project onto the space spanned by the column vectors of $M$ whose first $D$ entries correspond to vectors $w_{1,i},...,w_{N_i,i}$ and $\Pi Q_i \Pi$ wil have the desired property.

To choose $M$, suppose first that we are in the symmetric case.  Consider the projector onto the space orthogonal to the first $D$ rows of $M$.  This projector is symmetric, so it is a sum of symmetric rank-$1$ projectors of the form $|v_a\rangle\langle v_a|$, for real $v_a$, for $1\leq a \leq D'-D$.  Fill in the remaining $D'-D$ rows of $M$ with these vectors, letting the $D+a$-th row, for $1\leq a \leq D'-D$, contain entries correspond to the coordinate of the $v_a$.  In the self-dual case, proceed similarly, decomposing the projector into a sum of rank-$2$ projectors
$|v_a\rangle \langle v_a| + |w_a \rangle \langle w_a|$, where $w_a=Z \overline v_a$, for $1 \leq a \leq (D'-D)/2$.   Fill in the $D+2a$-th row with $w_b$ and fill in the $D+2a-1$-th row with $v_a$.
Choose the matrix $Z$ on the larger space to agree with the original $Z$ on the smaller space and to be
\be
\begin{pmatrix}
0 & 1\\
-1 & 0 &\\
&& 0 & 1\\
&& -1 & 0&\\
&&&&...
\end{pmatrix}
\ee
on the larger space.
\end{proof}
\end{lemma}

Finally, given that the $X'_i,Y'_i$ inherit the symmetry, the $V_i$ inherit the symmetry, and
lemma 8.4 in Ref.~\onlinecite{HL3} shows that the polar of a symmetric matrix is symmetric and the polar of a self-dual matrix is self-dual.

The symmetries manifest in a slightly different way in the case of vector bundles.  We say that $E$ is symmetric if
\be
\label{mapEs}
E(\vec \theta)=E(-\vec \theta)^T,
\ee
and we say that $E$ is self-dual if
\be
\label{mapEsd}
E(\vec \theta)=-Z' E(-\vec \theta)^T Z',
\ee
where the matrix $Z'$ again is a block matrix of the form
\be
\label{Zprimedef}
Z'=\begin{pmatrix} 0 & I \\ -I & 0 \end{pmatrix}.
\ee
We distinguish this matrix $Z'$ from the matrix $Z$ above the map $\um$ will increase the size of the matrices: if $E(\vec \theta)$ has a given size, then the matrices in $\um E$ are larger if $N>1$.
Eqs.~(\ref{mapEs},\ref{mapEsd}) correspond to real $K$-theory or twisted real $K$-theory.  We claim that the map $\um$ maps a symmetric or self-dual $E$ to a symmetric or self-dual local projector and the map $\vm$ maps a symmetric or self-dual local projector to a symmetric or self-dual $E$, respectively, for an appropriate choice of $Z,Z'$.

For the map $\vm$, we take $Z'=Z$ in the self-dual case.  Then, in either symmetric or self-dual case,
this property of $\vm$ can be shown as follows: we verify that if the local projector is symmetric or self-dual, then the strictly local Hamiltonian $H_{loc}$ has the same property.  Then, the twisted $H_{loc}(\vec \theta)$ does {\it not} have this property, but one can verify that $H_{loc}(\vec \theta)=H_{loc}(-\vec \theta)^T$ in the symmetric case (to verify this, note that complex conjugation of $H_{loc}(\vec \theta)$ changes the sign of $\vec \theta$ and recall that $H_{loc}(\vec \theta)$ is a Hermitian operator), or a similar property in the self-dual case.  Then, the projector onto the eigenspace of $H_{loc}(\vec \theta)$ with eigenvalue greater than $1/2$ inherits this property.

For the map $\um$, in the symmetric case, we write the matrices in the local projector in a basis which diagonalizes the $U_i$.  Then, the $U_i$ are symmetric by definition, and one may verify that the projector $P$ is symmetric.

The self-dual case of $\um$ is handled differently.  Now,
$Z'$ is given and it is necessary to define $Z$.   Let $Z'$ have dimension $2D'$, with basis vectors $v_1,....,v_{2D'}$ in the basis of Eq.~(\ref{Zprimedef}).  Then, for $1 \leq i \leq D$, to $v_{i+D}$, $Z'$ maps $v_i$, and it maps $v_{i+D}$ to $-v_i$.  Let $Z$ have dimension $2D'N^d$, with basis vectors $|\vec n\rangle \otimes v_i$.  Then, for $1 \leq i \leq D$, we define $Z$ to map $|\vec n\rangle \otimes v_i$ to $|-\vec n\rangle \otimes v_{i+D}$, and to map $\vec n \rangle \otimes v_{i+D}$ to $-|-\vec n\rangle \otimes v_i$.  That is, we change the sign of $\vec n$, in addition to mapping $v_i$ to $v_{i+D}$ or $v_{i+D}$ to $-v_i$.  The change of sign of $\vec n$ is defined mod $N$.
One may verify that for this choice of $Z$, the desired property holds.

The other seven cases are slightly more complicated.  In these case, the concept of the soft torus needs to modified, and similarly the map $\gm$ needs to be modified.  The reason is that the symmetries in these cases relate the eigenspace of $P$ with eigenvalue $1$ to the eigenspace of $P$ with eigenvalue $0$, and so that map $\gm$ as defined loses some of this information since it only considers one of the two space.  So, we deal with this case elsewhere.

\part{A Question About Quantum Channels}
In this part, we consider the ability of a
quantum channel to {\it simulate} another by means of suitable encoding
and decoding operations.
While classical channels have only two equivalence
classes under simulation (channels with non-vanishing capacity and
those with vanishing capacity), we show that there are a countable
infinity of different equivalence classes of quantum channels using
the example of the quantum erasure channel.  Specifically, we show
that an error channel with transmission probability $p=1/m$ cannot
be error corrected to any better transmission probability for integer
$m$.
We raise an open problem
regarding the ability to partially error correct a poor erasure channel:
for $p<1/2$ but $1/p$ non-integer, is it ever possible to improve the
transmission probability?
We present partial results in the direction that this is not possible,
but we do not succeed in
showing this more refined result.

Despite a range of rules of thumb for converting classical to quantum
information, such as one quantum bit being worth two classical bits as
in superdense coding\cite{superdense} or teleportation protocols\cite{teleport},
classical and quantum information are fundamentally
different.  In the absence of shared entanglement,
a classical channel is useless for transmitting quantum information.  No
matter how many times the classical channel is used, it cannot
transmit even a single qubit.
This is the sort of question we consider
here: given two resources, is it possible for one resource
to simulate the other, given arbitrarily many uses of the first
resource?  If not, then the two resources
are {\it qualitatively} different.

Shannon's noisy channel coding theorem\cite{shannon} implies that there
are only two qualitatively different kinds of classical channels, those
with non-vanishing capacity and those with vanishing capacity, because
given a noisy channel with non-vanishing capacity, error-correcting
decoding can be used
to transmit data
with arbitrarily small error probability, enabling
it to {\it simulate} any other classical channel,
as defined below.

There are at least three different kinds of quantum channels:
channels with non-vanishing quantum capacity, classical channels with non-zero capacity, and
classical channels with zero capacity.
However, there also exist channels with vanishing quantum capacity
which are still not classical channels.  A dramatic example of this,
and one of our motivations, is the discovery\cite{opnon}
that there exist pairs of quantum channels, both of which have vanishing
quantum capacity, but which can be used in tandem to transmit quantum
information.  In this letter we show that, under a precise
definition of simulation, there are infinitely many different
equivalence classes of quantum channels.

Similar ideas of simulation are developed in
the quantum resource theory\cite{qrt}.  The basic difference here
is that we ignore all quantitative differences, namely how many
uses of one channel are required to simulate another, and only ask
whether or not the simulation is possible.

\section{Definition of Simulation---}
We begin with some definitions.
Given two quantum channels, ${\cal C}, {\cal C'}$, we say that
${\cal C}$ $\epsilon$-simulates ${\cal C'}$ if there exists
an integer $n$, and quantum channels ${\cal E}$ and ${\cal D}$ such that
\be
\label{esimdef}
{\rm tr}(|{\cal D}({\cal C}^{\otimes n}({\cal E}(\rho)))-{\cal C}'(\rho)|)\leq \epsilon
\ee
for all density matrices $\rho$.  The norm used here is the trace norm.
The idea behind this definition is that by encoding the state $\rho$ using
the map ${\cal E}$, then transmitting over multiple uses of the channel
${\cal C}$, and then decoding with the map ${\cal D}$, we are able to
approximate the map ${\cal C}'$.
We say that 
${\cal C}$ simulates ${\cal C'}$ if ${\cal C}$ $\epsilon$-simulates
${\cal C}'$ for all $\epsilon>0$.  Note that if ${\cal C}$ simulates
${\cal C}'$, then ${\cal C}\otimes {\cal X}$ simulates ${\cal C}' \otimes
{\cal X}$ for any finite-dimensional channel ${\cal X}$.  Also,
simulation is transitive: if ${\cal C}$ simulates ${\cal C}'$ and
${\cal C}'$ simulates ${\cal C}''$, then ${\cal C}$
simulates ${\cal C}''$.

As an example of these definitions, let ${\cal C}$ and ${\cal C}'$
both be classical channels, in that the output of each channel depends only
on the diagonal elements of $\rho$ in some given basis.
Then, if ${\cal C}$ has
non-zero classical capacity, it can simulate any classical channel
${\cal C}'$.

Given a set of channels, such as the set of 
finite-dimensional classical
channels, we can define an equivalence relation on that set, such that
two channels ${\cal C},{\cal C}'$ are equivalent if and only if
${\cal C}$ can simulate ${\cal C}'$ and vice versa.  In the case
of the set of finite-dimensional classical channels, there are only two equivalence
classes: those with non-zero and those with zero capacity.

The set of finite-dimensional quantum channels includes at least
three different equivalence classes: the set of channels with vanishing
classical capacity, the set of classical channels with non-vanishing
classical capacity, and the set of channels with non-vanishing
quantum capacity.
However, there are clearly even more channels which are not within
any of these three equivalence classes.  For example,
a $50\%$ depolarizing channel has vanishing quantum capacity but can be
used to transmit quantum states using two-way communication\cite{bdsw} and hence
cannot be simulated by a classical channel.

\section{The Quantum Erasure Channel---}
The first result in this part is that there are a countable
infinity of different equivalence classes.  We consider
a quantum erasure channel which transmits a state with probability $p$.
Alice's input is a single qubit, and
Bob's output is a three dimensional space of
states, $|\uparrow\rangle,|\downarrow\rangle,|E\rangle$, with
\be
{\cal C}_p(\rho)=p \rho+(1-p)|E\rangle\langle E|,
\ee
where $|E\rangle$ indicates that the state is erased.
For $p>1/2$, the channel ${\cal C}_p$ has a non-vanishing quantum
capacity\cite{qerasurecode}, and so can simulate any quantum
channel.  Note that typically in the literature, the roles of
$p$ and $1-p$ are interchanged from our definition; we use
the above definition as it is more natural for what follows.

We will show that any erasure
channel ${\cal C}_p$ with $p=1/m$ for integer $m$,
erasure channel ${\cal C}_q$ with $q>p$.  Thus, for each $p=1/m$
the erasure channel ${\cal C}_p$ lies in a different
equivalence class.

The proof that ${\cal C}_p$ cannot simulate ${\cal C}_q$
for $p=1/m$ and $q>p$ follows from no-cloning.
Define a new channel ${\cal X}$ with one transmitter and $m$ receivers.
Each receiver receives the input state with probability $1/m$ and receives
an erasure flag otherwise.  That is,
\begin{eqnarray}
{\cal X}(\rho)&=&(1/m) \Bigl( \rho\otimes |E\rangle\langle E| \otimes ...\otimes |E\rangle\langle E| \\ \nonumber
&&+
|E\rangle\langle E| \otimes \rho \otimes |E\rangle\langle  E| \otimes ... \otimes |E\rangle\langle E|\\ \nonumber
&&+...\Bigr).
\end{eqnarray}
Suppose that for some $q>p$, for all $\epsilon>0$
there exist an integer $n$ and encoding and decoding channels ${\cal E}$ and ${\cal D}$
such that Eq.~(\ref{esimdef}) is satisfied.
Let Alice create an EPR pair, and then take half of the pair and encode it with
${\cal E}$ and send it through ${\cal X}$.  Let each receiver decode with ${\cal D}$.  
The idea of the proof is that there will be a non-zero probability of more than
one receiver decoding the state, in contradiction to no-cloning.

Consider first the case in which there exist an integer $n$ and channels ${\cal E},{\cal D}$
for which (\ref{esimdef}) holds with $\epsilon=0$.
Then since $q>1/m$
we have non-vanishing probability of being
in a state in which two different receivers share an EPR pair with Alice which
is impossible since no such state exists (we prove this more formally below).  

We now consider the case in which we have $\epsilon$-simulation for arbitrarily small
$\epsilon>0$, but not for $\epsilon=0$.
Let $\rho_{AB_i}$ denote the reduced density matrix of Alice and the $i$-th receiver, for
$i=1,...,N$, after Alice inputs half  of the EPR pair into the channel.
By the definition of $\epsilon$-simulation, for
each receiver, the reduced density matrix of Alice and that receiver is within
${\cal O}(\epsilon)$ of the state
\begin{eqnarray}
\label{rred}
&&(1-q) \Bigl(\frac{\openone}{2}\otimes |E\rangle\langle E|\Bigr) \\ \nonumber
&+& q (1/2) |\uparrow\downarrow-\downarrow\uparrow\rangle\langle\uparrow\downarrow-\downarrow\uparrow|.
\end{eqnarray}
Assuming it is possible to have $\epsilon$,
arbitrarily close to $0$, we can construct an infinite sequence of
multipartite states, $\rho^{\epsilon}$ such that
reduced density matrix
 $\rho_{AB_i}^{\epsilon}$ is within ${\cal O}(\epsilon)$ in trace
norm distance of the reduced density matrix in (\ref{rred}), with $\epsilon=1,1/2,1/3,1/4,...$  Since the
space of density matrices is compact, this sequence has a convergent
subsequence, which has a limit density matrix $\tilde \rho$ such
that for all $i$ $\tilde \rho_{AB_i}$ is exactly equal to Eq.~(\ref{rred}).

Given $\tilde \rho$, we can define a new matrix $\tilde \rho'$ which is obtained by having
each receiver measure whether or not they received an erasure flag.  That is,
define a string $s_1,...,s_n$ to be the {\it erasure pattern}.  Let $s_i=0$ indicate that the
$i$-th receiver received an erasure and $s_i=1$ indicate that the $i$-th receiver received
a state.  
Let $\Pi_0=|E\rangle\langle E|$ and
$\Pi_1=\openone-\Pi_0$.  
Let
\be
\Pi_{\vec s}=\Pi_{s_1} \otimes \Pi_{s_2} \otimes ... \otimes \Pi_{s_n}.
\ee
Then,
\be
\tilde \rho'=\sum_{\vec s} \Pi_{\vec s} \tilde \rho  \Pi_{\vec s},
\ee
where the sum is over strings $\vec s=(s_0,s_1,...,s_n)$.
Then, $\tilde \rho'$ is an incoherent sum of  density matrices $\Pi_{\vec s} \tilde \rho \Pi_{\vec s}$
with different erasure patterns.
However, since $q>1/m$, $\tilde \rho'$ has non-vanishing probability to have more than one
receiver share an EPR pair with Alice.  This means that for some string $\vec s$ which contains at least two $1$s in its erasure pattern,
the density matrices
$\Pi_{\vec s} \tilde \rho \Pi_{\vec s}$ must be non-vanishing,
in contradiction to no-cloning.

\section{The Case of Non-integer $1/p$}
We now turn to the case when $1/p$ is not an integer.  The question is, is it possible
for $p<1/2$ to do error correction to improve the transmission probability at all?
For example, would it possible possible to correct $C_{0.4}$ to $C_{0.401}$?
We have not been able to prove that this is impossible, but we have been able to prove that 
it is impossible for one (seemingly natural) type of decoder.

For any state $\rho$, the state ${\cal C}_p^{\otimes n}({\cal E}(\rho))$ is an incoherent sum of
different density matrices corresponding to different erasure patterns.  We have
\be
{\cal C}_p^{\otimes n}({\cal E}(\rho))=\sum_{\vec s} p^{\sum_i s_i} (1-p)^{\sum_i (1-s_i)} {\cal C}_{s_1} \otimes {\cal C}_{s_2}
\otimes ... \otimes {\cal C}_{s_n}({\cal E}(\rho)).
\ee
Note that
the density matrix ${\cal C}_{s_1} \otimes {\cal C}_{s_2}
\otimes ... \otimes {\cal C}_{s_n}({\cal E}(\rho)),
\rho_\{s_i\}$ is a density matrix in the $2^{\sum_i s_i}$-dimensional
space spanned by states which are qubits in the positions $j$ where $s_j=1$ and
erasures in the positions where $s_j=0$.
Define
\be
\rho_{\vec s}=
{\cal C}_{s_1} \otimes {\cal C}_{s_2}
\otimes ... \otimes {\cal C}_{s_n}({\cal E}(\openone/2)).
\ee
Define
\be
F(\vec s)=
1-\langle E|{\cal D}(
\rho_{\vec s}
)|E\rangle.
\ee
We call a decoder ${\cal D}$ deterministic if, for all strings $\vec s$, we have either
$F(\vec s)=0$ or
$F(\vec s)=1$.

Intuitively, a deterministic decoder is one which has the property that whether or not it output an erasure $|E\rangle$ is determined entirely by $\vec s$, namely the pattern of erasures it receives.

The second result in this part is that it is not possible for ${\cal C}_p$ to simulate ${\cal C}_q$ for $p\leq 1/2$ and  $q>p$ using
deterministic decoders.  That is, it is not the case that for every $\epsilon>0$ there exists
an
$n,{\cal E},{\cal D}$, with ${\cal D}$ a deterministic decoder, such  that (\ref{esimdef}) holds.
We prove this result in the next section.

The restriction to deterministic decoders seems natural.  However, there do exist encoders ${\cal E}$ for which the
best decoder is not deterministic.  For example, suppose Alice transmits a large ($n>>1$) number of qubits to Bob.
She transmits her state $\Psi$ in either the first or the second qubit she sends, picking one of the two at random,
and transmitting some fixed state, say $|\uparrow\rangle$, in the other.  She then uses the remaining
$n-2$ qubits to transmit the to Bob the {\it classical} information of which random choice she made.  Then, in order
for Bob to know if he can decode, he needs to know not just the erasure pattern (which of the first two qubits
were transmitted) but the contents of some of the remaining qubits to determine which of the first qubits to
use.  Certainly this encoding is not helpful to Alice and Bob in any way, but at least it demonstrates that
deterministic decoding is not possible for every encoding.

\subsection{Idea of Proof}
To understand the basic idea of the proof, consider first a simplified case in which whenever ${\cal D}$ outputs a qubit state then
it decodes the input state with perfect fidelity.  That is, if $F(\vec s)=1$, then
${\cal C}_{s_1} \otimes {\cal C}_{s_2}
\otimes ... \otimes {\cal C}_{s_n}({\cal E}(\rho)=\rho$ for all $\rho$.
Suppose further that $F$ has the following {\it monotonicity property} that
if $F(\vec s)=1$ for some string $\vec s$, then
if $F(\vec t)=1$ for all strings $\vec t$ such that $\vec t\geq \vec s$, where the inequality of strings
$\vec t\geq \vec s$ means that $t_i\geq s_i$ for all $i$.
This monotonicity property is also very natural: if it is possible to decode the qubit given a certain sequence of transmitted
qubits (described by the string $\vec s$), then it is also possible to decode if we also have access to additional qubits (those
qubits $i$ for which $\vec t_i=1$ and $\vec s_i=0$).

Define $f(p)$ to be the average of $F(\vec s)$ when the bits $s_1,...,s_n$ are chosen independently equal to $1$ with probability
$p$ and $0$ with probability $1-p$.
Then, if ${\cal C}_p$ simulates ${\cal C}_q$, we find that in the limit as $n\rightarrow \infty$
\be
f(p) \rightarrow q.
\ee
However, as we prove below, given that $f(p)>p$ and $p\leq 1/2$, it follows that $f(1-p)>1-p$.  
This allows us to prove a contradiction using no-cloning.
Consider a channel from Alice to two different receivers, Bob and Carol, where Alice's input is a qubit
and with probability $p$ Bob receives the qubit
Alice transmitted and Carol receives an erasure flag, and with probability $1-p$ Bob receives an erasure flag and Carol
receives the qubit Alice transmitted.  Then Carol can decode Alice's state with probability $f(1-p)$.
Thus, under the assumption that ${\cal C}_p$ simulates ${\cal C}_q$ with
a deterministic decoder, and under the further assumptions in this section, namely the property that ${\cal D}$ decodes
with perfect fidelity whenever it decodes and the monotonicity property, we find that with probability at least
$f(p)+f(1-p)-1$ both Bob and Carol receive Alice's state.  However, since $f(1-p)>1-p$, we have
$f(p)+f(1-p)-1>0$, violating no-cloning.

In the rest of the part we do two things to complete the proof of our second result, that
${\cal C}_p$ cannot simulate ${\cal C}_q$ for $p\leq 1/2$ and  $q>p$ using a deterministic decoder.
First, we remove the assumptions in
this section of perfect decoding and of monotonicity.  Second, we prove the claim that $f(p)>p$ for $p\leq 1/2$ implies that $f(1-p)>1-p$,
which we refer to as a ``Matthew principle"\cite{matthew}; this principle means that if error correction helps some channel with transmission
probability $p$, by making $f(p)>p$, it also helps a channel with transmission probability $1-p$.
This last claim is purely a statement in probability theory about the average of Boolean functions $F$ with the monotonicity property.

What we do in the next sections to deal with an imperfect decoder (one which does not decode with perfect fidelity, but
only with fidelity which tends to unity as $n$ tends to infinity) or a lack of monotonicity is to
construct a new decoding map
$\tilde {\cal D}$ that
Carol can use (with the {\it same} encoding map ${\cal E}$)
to $\epsilon'$-simulate a channel ${\cal C}_r$ with $r>1-p$, where
the quantity $\epsilon'$ tends to
zero as $\epsilon$ tends to zero.
That is, we prove that for any $\rho$,
\be
\label{holds}
{\rm tr}(|\tilde {\cal D}({\cal C}_{1-p}^{\otimes n}({\cal E}(\rho)))-{\cal C}_r(\rho)|)\leq \epsilon'.
\ee
The new decoding map $\tilde {\cal D}$ will be constructed to have the monotonicity property.
Given this, it is possible to
construct an infinite sequence of
tripartite states, $\rho_{ABC}^{\epsilon}$ such that the
reduced density matrix $\rho_{AB}^{\epsilon}$ is within $\epsilon$ in trace
norm distance of the reduced density matrix in (\ref{rred}) and the reduced
density matrix $\rho_{AC}^{\epsilon}$ is within $\epsilon'$ in trace norm of
\be
\label{rred2}
(1-r) (1/2) \openone\otimes |E\rangle\langle E| + r (1/2) |\uparrow\downarrow-\downarrow\uparrow\rangle\langle\uparrow\downarrow-\downarrow\uparrow|,
\ee
with $\epsilon=1,1/2,1/3,...$ and $\epsilon'$ tending to zero as $\epsilon$ tends to zero.
Since the
space of density matrices is compact, this sequence has a convergent
subsequence, which has a limit density matrix $\tilde \rho_{ABC}$ such
that $\tilde \rho_{AB}$ is exactly equal to Eq.~(\ref{rred}) and
similarly for $\tilde \rho_{AC}$ is equal to Eq.~(\ref{rred2}).  We claim that no such
$\tilde \rho_{ABC}$ can exist for $q+r>1$, as
for $q+r>1$
this density matrix $\tilde \rho_{ABC}$
must be non-zero when both Bob and Carol project
into the space orthogonal to $|E\rangle$.  Let $X$ be the positive
definite matrix that they obtain after projection.  Then $\sigma_{ABC}=X/{\rm Tr}(X)$
is a tripartite state, where
$\sigma_{AB}$ and $\sigma_{AC}$ both are EPR pairs.  Since this is not
possible, no such $\tilde \rho_{ABC}$ can exist for $q+r>1$, as claimed.

The only nontrivial step is to construct
the decoding map $\tilde {\cal D}$.
We make no attempt in this proof to optimize
constant factors, since we only care about the limit of
vanishing $\epsilon$.

\subsection{Imperfect Decoding}
We now show how to handle the situation in which the map ${\cal D}$ is deterministic but is not a perfect
decoder.  Note that if ${\cal D}$ is a perfect decoder but lacks the monotonicity property, then it
is easy to construct a new decoder, $\tilde {\cal D}$, that is a perfect decoder, has the monotonicity property, and
has a higher probability of decoding (that is, it gives a larger $f(p)$), as follows: if ${\cal D}$ decodes 
on a string $\vec s$, but not on a string $\vec t>\vec s$, then we define $\tilde {\cal D}$ to decode the string $\vec t$
simply by ignoring certain qubits.

We begin with some error estimates which will be useful to deal with the case in which ${\cal D}$ 
is an imperfect decoder.
Let $\vec s$ be a string for with $F(\vec s)=1$.
Define the ``average transmitted mistake rate" by
\be
E_{av}(\vec s)=
\int_{|\psi|=1} {\rm d}\psi \,
{\rm tr}\Bigl[
{\cal D}({\cal F}_{\vec s}(|\psi\rangle\langle\psi |))
\Bigl(1-|\psi\rangle\langle\psi|\Bigr)\Bigr].
\nonumber
\ee
That is, averaged over input pure states $|\psi\rangle$, this is
the probability that the output is orthogonal
to $|\psi\rangle$.

Define the ``maximum transmitted mistake rate" by
\be
\label{maxtrans}
E_{max}(\vec s)=
{\rm max}_{|\psi|=1}
{\rm tr}\Bigl[
{\cal D}({\cal F}_{\vec s}(|\psi\rangle\langle\psi |))
\Bigl(1-|\psi\rangle\langle\psi|\Bigr)\Bigr].
\ee
We now want to relate the averaged transmitted mistake rate to the
maximum transmitted mistake rate.
By relating the average
transmitted mistake rate to the maximum transmitted mistake rate, this will
allow us to bound the average (over different $\vec s$)
of the maximum transmitted mistake rate
of channels ${\cal F}_{\vec s}$.
We claim that
\be
\label{relation}
E_{av}(\vec s)\leq E_{max}(\vec s)\leq 4 E_{av}(\vec s).
\ee
The first inequality in Eq.~(\ref{relation}) is immediate.  To show
the second inequality, let $\psi$ be the vector which maximizes
(\ref{maxtrans}).  Let $\psi^\perp$ be a vector orthogonal to $\psi$.
Then, we can write any vector as $\cos(\theta) z \psi+\sin(\theta) w \phi$,
where $z,w$ are phases ($|z|=|w|=1$) and $\theta$ is an angle.
The transmitted mistake rate is a quartic function of the vector,
and so must be of the form $a \cos(\theta)^4+b\cos(\theta)^2\sin(\theta)^2+
c\sin(\theta)^4$ after averaging over $z,w$, where $a=E_{\max}(\vec s)$.  
The average of this function over angle is $a/3+b/6+c/3$.  Since the
function is positive, we have $a,c\geq 0$ and $|b|\leq 2\sqrt{ac}$.  Minimizing 
$a/3-\sqrt{ac}/3+c/3$ over $c$, we find that the average is at least
$a/4$.  It is possible that there exist tighter bounds relating the
maximum transmitted mistake rate to the average.

Note that
\be
\label{pfrompt}
\sum_{\vec s} 
p^{\sum_i s_i} (1-p)^{\sum_i (1-s_i)} F(\vec s) \geq q-\epsilon.
\ee
By Eq.~(\ref{esimdef}),
\begin{eqnarray}
\label{claimthat}
\sum_{\vec s} 
p^{\sum_i s_i} (1-p)^{\sum_i (1-s_i)}
F(\vec s) E_{av}(\vec s)
&\leq & \\ \nonumber
\int_{|\psi|=1} {\rm d}\psi \,
{\rm tr}\Bigl[
\Pi_1
{\cal D}({\cal C}^{\otimes n}({\cal E}(|\psi\rangle\langle\psi |)))
\Pi_1
\Bigl(1-|\psi\rangle\langle\psi|\Bigr)\Bigr]
&\leq &\epsilon.
\end{eqnarray}

Write the decoding map ${\cal D}$ as
\be
\nonumber
{\cal D}=\sum_{\vec s} {\cal D}_{\vec s},
\ee where
${\cal D}_{\vec s}(\rho)={\cal D}(\Pi_{\vec s}\rho \Pi_{\vec s})$.  That is, ${\cal D}_{\vec s}$ is the decoding map that
is used when a given erasure pattern $\vec s$ occurs.
Define a map ${\cal D}'$ as follows.  
For each string $\vec s$, if
$E_{av}(\vec s)\leq \sqrt{\epsilon}$, set ${\cal D}'_{\vec s}(\rho)={\cal D}_{\vec s}(\rho)$.
On the other hand, if 
$E_{av}(\vec s)>\sqrt{\epsilon}$, we set ${\cal D}'_{\vec s}(\rho)=|E\rangle\langle E| {\rm tr}(\rho)$.
We then define
\be
{\cal D}'(\rho)=\sum_{\vec s} {\cal D}'_{\vec s}(\Pi_{\vec s} \rho \Pi_{\vec s}).
\nonumber
\ee

We now construct the new decoding map $\tilde {\cal D}$ which has the monotonicity property; the idea is that if
it is possible to
accurately decode a message when certain of the qubits
are transmitted perfectly, it must also be possible to decode the message
when
a superset of those bits are transmitted perfectly.  
We first
define ${\cal H}_{\vec u}$ to be a map which ``hides" data from the decoder
as follows:
\be
{\cal H}_{\vec u}(\rho)={\cal C}_{u_1}\otimes {\cal C}_{u_2} \otimes ...
\otimes {\cal C}_{u_n}(\rho).
\nonumber
\ee
That is, wherever $\vec u$ has a zero entry, the map $H$ produces an error
output, but otherwise it transmits perfectly.
Then, to define $\tilde {\cal D}$, for any $\vec s$, consider the set of all strings $\vec t$, such
that $t_i \leq s_i$ for all $i$, and such that $E_{av}(\vec t)\leq \epsilon_0$,
so
that ${\cal D}'_{\vec s}$ outputs a state in the qubit subspace, rather than in the error
subspace.
If this set is non-empty, let $\vec t_{max}$ denote the string in this set
which has the smallest average transmitted mistake rate and let
\be
\tilde {\cal D}_{\vec s}(\rho)={\cal D}'_{\vec t_{max}}({\cal H}_{\vec t_{max}}(\rho)).
\nonumber
\ee
If this set is empty, let $\tilde {\cal D}_{\vec s}(\rho)={\rm tr}(\rho)|E\rangle\langle E|$.

Let $\tF(\vec s)=1$ if $\tilde {\cal D}$ outputs a state in the qubit subspace when erasure
pattern $\vec s$ occurs, and let $\tF(\vec s)=0$ if $\tilde {\cal D}$ outputs an erasure.
Note that $\tF(\vec s)\geq F(\vec s)$.

\subsection{$\epsilon'$-simulation for Carol}
The average transmission
probability, $r$, that Carol obtains using $\tilde {\cal D}$
is equal to
\be
r=\sum_{\vec s\in T}
(1-p)^{\sum_i s_i} p^{\sum_i (1-s_i)}.
\nonumber
\ee

From Eq.~(\ref{relation}) and the definition of ${\cal D}'$,
it follows that Eq.~(\ref{holds}) holds for
$\epsilon'=\sqrt{4} \epsilon^{1/4}+{\cal O}(\epsilon^{3/8})$.  The
only remaining step is to show that, for sufficiently small $\epsilon$,
\be
\label{only}
r>1-p.
\ee

\subsection{Average Transmission Probability}
Eq.~(\ref{only}) corresponds to
the following problem in classical probability.  We have $n$ bits, each
chosen independently to equal $1$ with probability $p$ and $0$ with
probability $1-p$.
We have a function $\tF(\vec s)$ which obeys the monotonicity property: $\vec t>\vec s\rightarrow
\tF(\vec t)\geq \tF(\vec s)$.
Such an $\tF$ defines a function $\tf(p)$, which is the probability that $\tF(\vec s)=1$; $\tf(p)$ is the sum
\be
\tf(p)=\sum_{\vec s \in T} 
p^{\sum_i s_i} (1-p)^{\sum_i (1-s_i)}.
\nonumber
\ee
By the choice of $\epsilon_0$, and by Eq.~(\ref{claimthat}),
$\tf(p)\geq
\sum_{\vec s} p^{\sum_i s_i} (1-p)^{\sum_i (1-s_i)} F(\vec s)-\sqrt{\epsilon}$, so by
Eq.~(\ref{pfrompt}),
\be
\tf(p)\geq q-\epsilon-\sqrt{\epsilon},
\ee
and so $\tf(p)>p$ for sufficiently small $\epsilon$.

\begin{figure}
\centerline{
\includegraphics[scale=0.2]{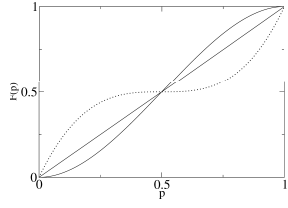}}
\caption{$\tf(p)$ can equal $p$ (if $\tF(\vec s)=s_1$, for example),
or $\tf(p)$ can start smaller than $p$ and increase
to greater than $\tf(p)$ (we plot the case
$(s_1 \wedge s_2) \vee (s_2 \wedge s_3) \vee (s_1 \wedge s_3)$), but
the dotted curve is not possible by the Matthew principle for $\tf(p)$.}
\label{figqc}
\vspace{5mm}
\end{figure}

We will show that if $\tf(p)>p$ for some $p<1/2$, then $\tf(p)+\tf(1-p)>1$; since
$f(p)>p$ and $\tF\geq F$, it follows that $\tf(p)>p$.
We will use induction and an unfortunate amount of algebra to show that
for any $\tF(\vec s)$ with the monotonicity property
at least one of the following three cases is
true: (1) $\tf(p)=0$ for all $p$; or (2) $\tf(p)=1$ for all $p$;
or (3) 
\begin{eqnarray}
\nonumber
\partial_{p} \ln[g(p)]& \geq & \partial_p \ln[p/(1-p)] \\ \nonumber
&=& 1/[p(1-p)],
\end{eqnarray}
where
$g(p)$ is the success-to-failure ratio:
\be
g(p)=\frac{\tf(p)}{1-\tf(p)}.
\nonumber
\ee
$g(p)$ is a monotonically increasing function of $p$.
This will prove the desired result, since then if $\tf(p)>p$, either $\tF(1-p)=1$ (case 2) or $\tF(1-p)>1-p$ by case 3.
As shown in Fig.~\ref{figqc}, this
result may be regarded as a kind of ``Matthew principle" for
error correction: roughly speaking, if error correction helps a channel
for some $p$, it also helps a channel with transmission
probability $p'>p$.

To show the result,
consider the $n$-th bit.  Let $\tf_{1}(p)$ denote the expectation
value of $\tf(p)$ conditioned on the $n$-th bit being equal to one, and let $\tf_{0}(p)$ denote the
expectation value of $\tf(p)$ conditioned on the $n$-th bit being equal to zero.
Note that $\tf_{1}(p)\geq \tf_{0}(p)$.
Define $G_{s}(p)=\tf_s(p)/[1-\tf_s(p)]$ for $s=0,1$.
Consider the case in which $0<\tf_0(p)\leq \tf_1(p)<1$ (the remaining cases are
simpler).  We have
$\partial_p \ln[\tf(p)]=[\tf_1-\tf_0+p\partial_p \tf_1(p)+(1-p) \partial_p \tf_0(p)]/\tf(p)$,
and similarly for $\partial_p \ln[1-\tf(p)]$, giving
\be
\label{inint}
\partial_p \ln[g(p)]=\frac{\tf_1-\tf_0+p\partial_p \tf_1(p)+(1-p) \partial_p \tf_0(p)}{\tf(p)[1-\tf(p)]}.
\ee
The induction hypothesis tells us that $\partial_p \tf_s(p)\geq \tf_s(p)[1-\tf_s(p)]/[p(1-p)]$.
Inserting the induction hypothesis into Eq.~(\ref{inint}) and multiplying
by $p(1-p)$ gives
$p(1-p) \partial_p \ln[g(p)]
\geq \frac{p(1-p)(\tf_1-\tf_0)+p \tf_1(p) [1-\tf_1(p)] +(1-p) \tf_0(p)[1-\tf_0(p)]}{\tf(p)[1-\tf(p)]}$.
We wish to show that the right-hand side of this inequality is greater than
or equal to unity, or equivalently, we wish to show that
\begin{eqnarray}
\label{inint2}
&&
p(1-p)(\tf_1-\tf_0)+p \tf_1(p) [1-\tf_1(p)] \\ \nonumber
&& +(1-p) \tf_0(p)[1-\tf_0(p)]
\\ \nonumber
&\geq & \tilde f(p)[1-\tilde f(p)]
\\ \nonumber
&=& [p \tf_1(p)+(1-p) \tf_0(p)][p (1-\tf_1(p))+(1-p) (1-\tf_0(p))].
\end{eqnarray}
After subtracting $p^2 \tf_1(p) (1-\tf_1)+(1-p)^2 \tf_0 (1-\tf_0(p))$ from both
sides, and dividing by $p(1-p)$, Eq.~(\ref{inint2}) becomes
$\tf_1-\tf_0+\tf_1-\tf_1^2+\tf_0-\tf_0^2\geq \tf_0+\tf_1-2 \tf_1 \tf_0$,
which is equivalent to $\tf_1-\tf_0\geq (\tf_1-\tf_0)^2$, which is true because
$0\leq \tf_0 \leq \tf_1 \leq 1$.  This completes the proof.

\section{Discussion}
Despite ignoring all {\it quantitative} details of the relation between channels, we have found a rich structure, with
infinitely
many different
equivalence classes.
There exist pairs of channels that cannot simulate each other.
For example, the operational non-additivity of the pair of channels considered in
\cite{opnon} implies that that this is a pair of channels which cannot simulate each other.
The full structure of equivalence classes
of quantum channels, and which classes can simulate other classes, promises
to be very complicated.
We leave as an open problem the question of whether it is possible for ${\cal C}_p$ to simulate ${\cal C}_q$ for
$p\leq 1/2$ and $q>p$ using non-deterministic decoders.


\begin{thebibliography}{99}
\bibitem{mporev} D. Perez-Garcia, F. Verstraete, M. M. Wolf, and J. I. Cirac, ``Matrix Product State Representations", QIC {\bf 7}, 401 (2007).

\bibitem{aklt} I. Affleck, T. Kennedy, E. H. Lieb and H. Tasaki, Commun. Math. Phys. {\bf 115}, 477 (1988).

\bibitem{fnw} M. Fannes, B. Nachtergaele and R. F. Werner, ``Finitely Correlated States on Quantum Spin Chains", Commun. Math. Phys. {\bf 144}, 443-490 (1992).

\bibitem{areaexp} F. G. S. L. Brandao and M. Horodecki, ``Exponential Decay of Correlations Implies Area Law ", arXiv:1206.2947.

\bibitem{qexp1} M. B. Hastings, ``Entropy and Entanglement in Quantum Ground States", Phys. Rev. B. {\bf 76}, 035114 (2007).

\bibitem{qexp2} A. Ben-Aroya and A. Ta-Shma, ``Quantum expanders and the quantum entropy difference problem", arXiv:quant-ph/0702129.

\bibitem{rugqe} M. B. Hastings, ``Random Unitaries Give Quantum Expanders",  Phys. Rev. A {\bf 76}, 032315 (2007). 

\bibitem{peps} F. Verstraete and J. I. Cirac, ``Renormalization algorithms for Quantum-Many Body Systems in two and higher dimensions", arXiv:cond-mat/0407066.

\bibitem{read} J. Dubail and N. Read, ``Tensor network trial states for chiral topological phases in two dimensions", arXiv:1307.7726.

\bibitem{peps1} Z.-C. Gu, F. Verstraete and X.-G. Wen, arXiv:1004.2563.

\bibitem{peps2} B. B\'{e}ri and N. Cooper, Phys. Rev. Lett. 106, 156401
(2011).

\bibitem{hasmich} M. B. Hastings and S. Michalakis, arXiv:1306.1258.

\bibitem{dpg} D. P\'{e}r\'{e}z-Garcia, M. Sanz, C. E. Gonzalez-Guillen, M. M. Wolf, and J. I. Cirac,
``A canonical form for Projected Entangled Pair States and applications", New J. Phys. {\bf 12}, 025010 (2010).


\bibitem{voiculescu} D. Voiculescu, Acta Sci. Math {\bf 45}, 429–431, (1983).

\bibitem{loring} R. Exel and T. Loring, ``Almost Commuting Unitary Matrices", Proc. Amer. Math. Soc., {\bf 106}, 913-915 (1989); R. Exel. and T. A. Loring, ``Invariants of almost commuting unitaries", J. Funct. Anal., {\bf 95}, 364 (1991).

\bibitem{loring2} T. A. Loring and A. P. W. S\o{}rensen, arXiv:1107.4187.

\bibitem{HL1} M. B. Hastings and T. A. Loring,
``Almost commuting matrices, localized Wannier functions, and the quantum Hall effect",
J. Math. Phys. {\bf 51}, 015214 (2010).

\bibitem{HL2} T. A. Loring and M. B. Hastings,
``Disordered Topological Insulators via C$^*$ Algebras",
Europhysics Lett. {\bf 92}, 67004 (2010).

\bibitem{HL3} M. B. Hastings and T. A. Loring, ``Topological Insulators and $C^*$-Algebras: Theory and Numerical Practice", Ann. Phys. {\bf 326}, 1699 (2011).

\bibitem{POVM} A POVM is a standard concept in quantum information theory.  A POVM is a set of positive semi-definite operators $E_i$, indexed by some index $i$, such that $\sum_i E_i=I$.

\bibitem{Naimark} M. A. Naimark, ``Spectral functions of a symmetric operator", Izv. Akad. Nauk. SSR Ser. Mat. {\bf 4}, 277-318 (1940).

\bibitem{fluxtorus} The term ``flux torus" is used for a space of gauge fields modulo pure gauge transformations.  Our choice of how to twist boundary conditions corresponds to a particular choice of gauge.  See J. E. Avron, ``Adiabatic Quantum Transport", Les Houches Session LXI 1994,
``Mesoscopic Quantum Physics", E. Akkermans and G. Montambaux, J. L. Pichard and J. Zinn-Justin, Eds, North-Holand (1995) and
D. Thouless, M. Kohmoto, M. Nightingale, and M. den Nijs, Phys. Rev. Lett. {\bf 49}, 405 (1982).

\bibitem{kitaev} A. Kitaev, ``Periodic table for topological insulators and superconductors", in Proceedings of the L.D.Landau Memorial Conference "Advances in Theoretical Physics", June 22-26, 2008, Chernogolovka, Moscow region.

\bibitem{topo10} A. P. Schnyder, S., A. Furusaki, A. W. W. Ludwig , ``Classification of topological insulators and superconductors in three spatial dimensions",
Phys. Rev. B {\bf 78}, 195125 (2008) .

\bibitem{10rmt} M. R. Zirnbauer, J. Math. Phys. {\bf 37}, 4986 (1996);
A. Altland and M. R. Zirnbauer, Phys. Rev. B {\bf 55}, 1142 (1997).

\bibitem{superdense} C. H. Bennett and S. J. Wiesner, Phys. Rev. Lett.
{\bf 69}, 2881 (1992).

\bibitem{teleport} C. H. Bennett, G. Brassard, C. Cr\'{e}peau,
R. Jozsa, A. Peres, and W. K. Wotters, Phys. Rev. Lett. {\bf 70},
1895 (2003).

\bibitem{shannon}  C. E. Shannon, Bell Syst. Tech. Jour. {\bf 27}, 379 (1948).

\bibitem{qrt} I. Devetak, A. W. Harrow, and A. Winter, IEEE
Trans. Inf. Th. {\bf 54}, 4587 (2008).

\bibitem{bdsw} C. H. Bennett, D. P. DiVincenzo, J. A. Smolin, and W. K.
Wootters, Phys. Rev. A {\bf 54}, 3824 (1996).

\bibitem{opnon} G. Smith and J. Yard, Science {\bf 321}, 1812 (2008).

\bibitem{qerasurecode} C. H. Bennett, D. P. DiVincenzo, and J. A.
Smolin, Phys. Rev. Lett. {\bf 78}, 3217 (1997).

\bibitem{matthew} This principle refers to giving more to those that
already have something, and has nothing to do with the first name of the
author.


\end{thebibliography}
\end{document}